\documentclass{aa}  
\usepackage{graphicx}
\usepackage{txfonts}
\usepackage{epsfig}
\usepackage[below]{placeins}
\usepackage{natbib}
\usepackage{journal_shortcuts}
\bibpunct{(}{)}{;}{a}{}{,}
\begin{document}
   \title{Metal-rich multi-phase gas in M87}

   \subtitle{AGN-driven metal transport, magnetic-field supported multi-temperature gas, and constraints on non-thermal emission observed with XMM-Newton}
   \author{A. Simionescu \inst{1}
          \and N. Werner \inst{2,3}
	  \and A. Finoguenov \inst{1,4}
	  \and H. B\"ohringer \inst{1}
	  \and M. Br\"uggen \inst{5}
          }

   \offprints{A. Simionescu, aurora@mpe.mpg.de}

   \institute{Max Planck Institute for Extraterrestial Physics,
              Giessenbachstr, 85748 Garching, Germany
        \and
	 SRON Netherlands Institute for Space Research, Sorbonnelaan 2, NL - 3584 CA Utrecht, the Netherlands
         \and
             Max Planck Institute for Astrophysics,
	     Schwarzschildstr 1, 85748 Garching, Germany
	\and
	University of Maryland Baltimore County, 1000 Hilltop Circle, Baltimore, MD, 21250, USA
	\and
	Jacobs University Bremen, P.O. Box 750 561, 28725 Bremen, Germany
             }

   \date{Received ; accepted }

% \abstract{}{}{}{}{}

  \abstract
  {
We use deep ($\sim$120~ks) XMM-Newton data of the M87 halo to analyze its spatially resolved temperature structure and chemical composition. 
%We perform spectral fitting on data extracted from spatial regions with S/N of 200. 
We focus particularly on the regions of enhanced X-ray brightness associated with the inner radio lobes, which are known not to be described very well by single-temperature spectral models. Compared to a simple two-temperature fit, we obtain a better and more physical description of the spectra using a model that involves a continuous range of temperatures in each spatial bin. The range of temperatures of the multiphase gas spans $\sim$0.6--3.2 keV. Such a multiphase structure is only possible if thermal conduction is suppressed by magnetic fields.
In the multi-temperature regions, we find a correlation between the amount of cool gas (with a temperature below that of the surrounding X-ray plasma) and the metallicity, and conclude that the cool gas is more metal-rich than the ambient halo. In the frame of the assumed thermal model, we estimate the average Fe abundance of the cool gas to $\sim$2.2 solar. Our results thus point toward the key role of the active galactic nucleus (AGN) in transporting heavy elements into the intracluster medium by uplifting cool, metal-rich gas from the galaxy. However, the abundance ratios of O/Si/S/Fe in and outside the X-ray arms are similar, indicating that the dominant fraction of metals in the gas halo was uplifted by AGN outbursts relatively recently compared to the age of M87. 
Our best estimate for the mass of the cool gas is $5 \times 10^8 M_\odot$, which probably stems from a mixture of ICM, stellar mass loss, and Type Ia supernova products. $\approx$30--110 Myr are required to produce the observed metals in the cool gas. 
Finally, we put upper limits on possible non-thermal X-ray emission from M87 and, combining it with the 90~cm radio maps, we put lower limits of around $\sim$0.5--1.0~$\mu$G on the magnetic field strength.

}
   \maketitle
%
%________________________________________________________________

\section{Introduction}
%_________________________________________________________________

The impact of active galactic nuclei (AGN) is currently the most promising solution for the cooling flow problem in galaxy clusters: e.g., \citet{Binney95,Brueggen02,Boehringer02,Birzan04,Omma04,Sijacki07} and references therein; for a review see \citet{Peterson06,McNamara07}. It is moreover invoked to transport heavy elements into the intracluster medium (ICM) \citep{Brueggen02metals,Omma04,Rebusco06,Roediger06} and to explain the exponential cut-off of the bright end of the galaxy luminosity function \citep[e.g.][]{Croton06,Sijacki07}. 

The AGN-ICM interaction is one of the driving motivations for deep observations that allow detailed analysis of the properties and spatial structure of the intracluster gas. While for a complete understanding of the ICM physics a combined analysis of multiwavelength data is usually necessary, most of the information about the ICM is gained from observations in the X-ray domain in which this hot plasma emits most of its radiation. The main targets of such detailed analyses are thus in particular nearby X-ray bright objects, where both good spatial resolution and high spectral statistics can be achieved. 

This makes the hot gas halo of M87, which is the second brightest extragalactic X-ray source in the sky located at the center of the nearby Virgo cluster at a distance of only 16 Mpc \citep{Tonry01}, an ideal target for studying the AGN-ICM interaction in great detail. M87 is known to host an AGN powered by the galaxy's central supermassive black hole with a mass of $3.2\times10^9M_{\odot}$ \citep{Harms94}.  The AGN jet and unseen counterjet are believed to drive the complex large-scale system of lobes observed in the radio domain \citep[e.g.][]{OEK00}, which clearly interact with the X-ray gas. The evidence for this is seen primarily in the spatial correlation between the two ``inner'' radio lobes extending east (E) and southwest (SW) of the core of M87 and regions of increased X-ray surface brightness, also referred to as the E and SW X-ray arms \citep{Feigelson87, Boehringer95, Belsole01,Young02, Forman05, Forman06}. 

The impact of the AGN on the X-ray emitting gas in M87 is however not limited to the surface brightness enhancement. Also the spectral properties in the E and SW arms are affected, these regions being the only part of the M87 halo which require more complex multi-temperature spectral modeling while the rest of the gas is well approximated by single-temperature models \citep{Molendi02,Matsushita02,Belsole01}. A possible explanation for this is given by \citet{Churazov01}, who suggest that the radio lobes rise buoyantly through the hot plasma, uplifting cooler gas from the central region and mixing it with ambient gas at larger radii. If this model is correct, then M87 offers a unique chance to study the effects of AGN-induced gas mixing and gas transport and the influence of these phenomena on the energy balance in the cooling core and on the metal distribution in the gas halo.

In this paper we use deep XMM-Newton X-ray observations of the M87 halo to characterize and understand the temperature structure and spatial variation of the multi-temperature gas in the X-ray arms, and to investigate its correlation with the radio plasma. Moreover, we aim to map the distribution of various metals and verify the influence of the AGN and the radio lobes in distributing and transporting these heavy elements from the galaxy center into the M87 hot gas halo.

The paper is laid out as follows: in Sect. \ref{sect:obs_data_analysis} we present the data sets and data reduction techniques employed; in Sect. \ref{sec:model} we describe the various multi-temperature spectral models used in our analysis; in Sect. \ref{sec:tstr} we discuss the thermal structure and spatial distribution of the multi-temperature phases and comment on possible suppression of the Spitzer heat conduction; in Sect. \ref{sect:spatial_distr_metals} we present the spatial distribution of the metals in the M87 gas halo and the abundance patterns of different metals relative to Fe in and outside the radiolobe regions; in Sect. \ref{sect:coolgas} we estimate and discuss the mass and metallicity of cool gas uplifted by the AGN; finally, in Sect. \ref{sect:pow} we give an upper limit for non-thermal emission from the radiolobe regions. Our conclusions are summarized in Sect. \ref{sect:conclusions}. We adopt a redshift for M87 of z=0.00436, a luminosity distance of 16 Mpc \citep{Tonry01}, and a scale of 4.65 kpc per arcminute.
 
%________________________________________________________________

\section{Observations and data analysis}\label{sect:obs_data_analysis}
%_________________________________________________________________

M87 was observed with XMM-Newton for 60 kiloseconds (ks) on June 19, 2000 and re-observed for 109 ks on January 10, 2005. After removing the periods affected by soft-proton flares, the net total exposure time is 85 ks for EPIC/pn and 120 ks for each EPIC/MOS detector. This deep exposure of such a near and bright source allows us to analyze the gas halo around M87 in great detail and with excellent statistics.

For the analysis, we used the 7.0.0 version of the XMM-Newton Science Analysis System (SAS) and employed the standard analysis methods as described in e.g. \cite{Watson01}. Point sources were found using the source detection algorithm implemented in SAS and removed from the observation after a visual check to eliminate spurious detections. The spectra from each of the two PN observations were corrected for out-of-time events separately. We note that our observations were affected by gain calibration uncertainties, in some cases the spectra - especially for the PN detector - being blue-shifted by up to $\sim$1500 km/s. The MOS detectors were also influenced by this shift, but the effect was a factor of $\sim$3 less strong than for the PN. To account for this, we artificially blue-shifted our spectral models.

As background we used a collection of blank-sky maps \citep{ReadPonman} scaled according to the corresponding exposure times for each detector and each observation. The accuracy of this scaling for the present dataset is discussed in \cite{Simionescu07}.

We used an adaptive binning method based on weighted Voronoi tessellations \citep{Voronoi06}, which is a generalization of the algorithm presented in \cite{Cappellari03}, to bin the combined counts image to a signal-to-noise ratio (SNR) of 200 in the energy range 0.5-2.0 keV, corresponding to 40000 counts per spatial bin. The binning was constrained to follow four different radio contours of the 90 cm radio image at 10, 3, 1 and 0.3 mJy (each region between two successive radio contours was binned separately). The region inside the highest chosen radio contour was not fitted in order to avoid possible contamination by the AGN, whose flux was particularly high in the second observation causing photon pileup in the EPIC detectors \citep{Werner06}.
The spectra were fitted in the energy range between 0.4-7.0 keV with the various models described in the next section. We fitted the spectra from all six datasets (two observations with three detectors each) simultaneously with the same model parameters except for a factor to account for variations in the relative overall normalization between the detectors. For this, the PN of the new observation was taken as reference.

%________________________________________________________________

\section{Spectral models}\label{sec:model}
%_________________________________________________________________

For the spectral analysis we use the SPEX spectral fitting package \citep{spex}. We fit the observed spectra with two models: a
simple two-temperature MEKAL model in which the abundances of the two thermal components are coupled to each other, and a differential emission
measure (DEM) model with a cut-off power-law distribution of emission measures versus temperature ({\it{wdem}}). 

The emission measure $Y = \int n_{\mathrm{e}} n_{\mathrm{i}} dV$ (where $n_{\mathrm{e}}$ and $n_{\mathrm{i}}$ are the electron and ion densities, $V$ is the volume of the emitting region) in the {\it{wdem}} model is specified in Eq.~(\ref{eq:wdem}) following from \citet{Kaastra04}:
\begin{equation}
\frac{dY}{dT} = \left\{ \begin{array}{ll}
AT^{1/\alpha} & \hspace{1.0cm} T_{\mathrm{min}} < T < T_{\mathrm{max}}, \\
0 & \hspace{1.0cm} \mathrm{elsewhere}. \\
\end{array} \right.
\label{eq:wdem}
\end{equation}

The emission measure distribution has a cut-off at $T_{\mathrm{min}}=cT_{\mathrm{max}}$.
For $\alpha \to \infty$ we obtain a flat emission measure distribution. 

The fraction of the emission measure of the gas lying below a chosen temperature $T_{\mathrm{maxcool}}$ to the total emission measure is given by:
\begin{equation}
\frac{Y_{\mathrm{maxcool}}}{Y}=\frac{\int^{T_{\mathrm{maxcool}}}_{T_{\mathrm{min}}} \frac{dY}{dT}\,dT}{\int^{T_{\mathrm{max}}}_{T_{\mathrm{min}}} \frac{dY}{dT}dT}.
\end{equation}
By integrating this equation we obtain a direct relation for $Y_{\mathrm{maxcool}}/Y$ as a function of $\alpha$ and $c$:
\begin{equation}
\frac{Y_{\mathrm{maxcool}}}{Y}=\frac{{(T_{\mathrm{maxcool}}/T_{\mathrm{max}})}^{1/\alpha+1}-c^{1/\alpha+1}}{1-c^{1/\alpha+1}}.
\label{eq:ycool}
\end{equation}

The {\it{wdem}} model appears to be a good empirical approximation for the spectrum in cooling cores of clusters of galaxies \citep[e.g.][]{Kaastra04,Werner06_2A0335,dePlaa06}. 

In the spatial bins where either the normalization of the cool component for the two-temperature fit or the lower cutoff for the {\it{wdem}} model were less than $2\sigma$ significant, we fixed the temperature of the cool component to 1 keV and the lower cutoff to $0.3 T_{\mathrm{max}}$, respectively. Unless otherwise stated we assume a hydrogen column density of $2.0 \times 10^{20} {\mathrm{cm}}^{-2}$ \citep{Lieu96}.

Throughout the paper we give the measured abundances relative to the proto-solar values given by \citet{Lodders}. The recent solar abundance determinations by \citet{Lodders} give significantly lower abundances of oxygen, neon and iron than those measured by \citet{Angr}.

%________________________________________________________________

\section{Temperature structure of the M87 halo}\label{sec:tstr}
%_________________________________________________________________

\subsection{Two-temperature models}

\begin{figure}[tbp]
\begin{center}
\includegraphics[width=\columnwidth]{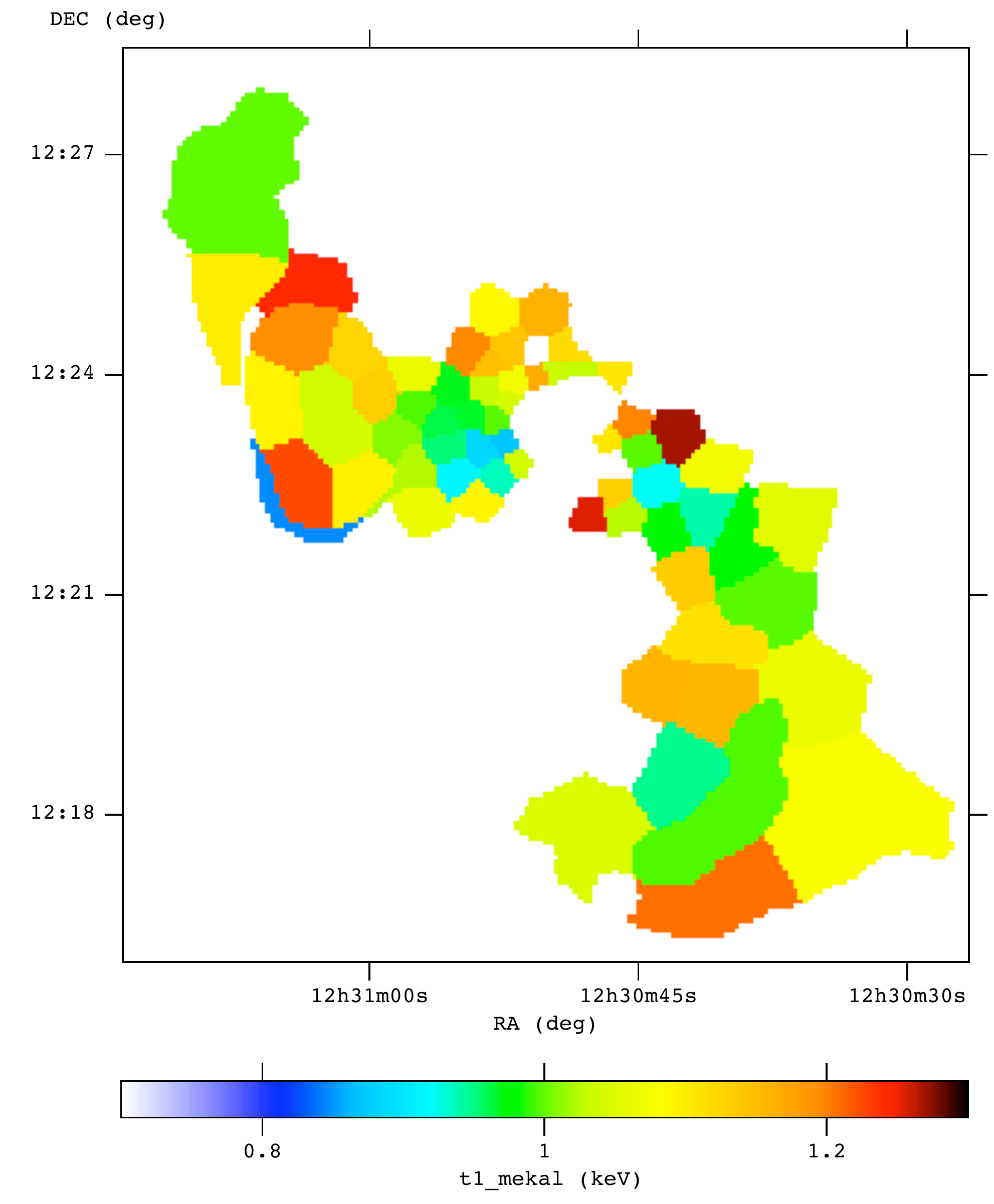}\\
\caption{Temperature of the cool component as determined from the two-temperature MEKAL fit. Only the bins where the normalization of the cool component was more than $3 \sigma$ significant are shown.}
\label{t1_spex}
\includegraphics[width=\columnwidth]{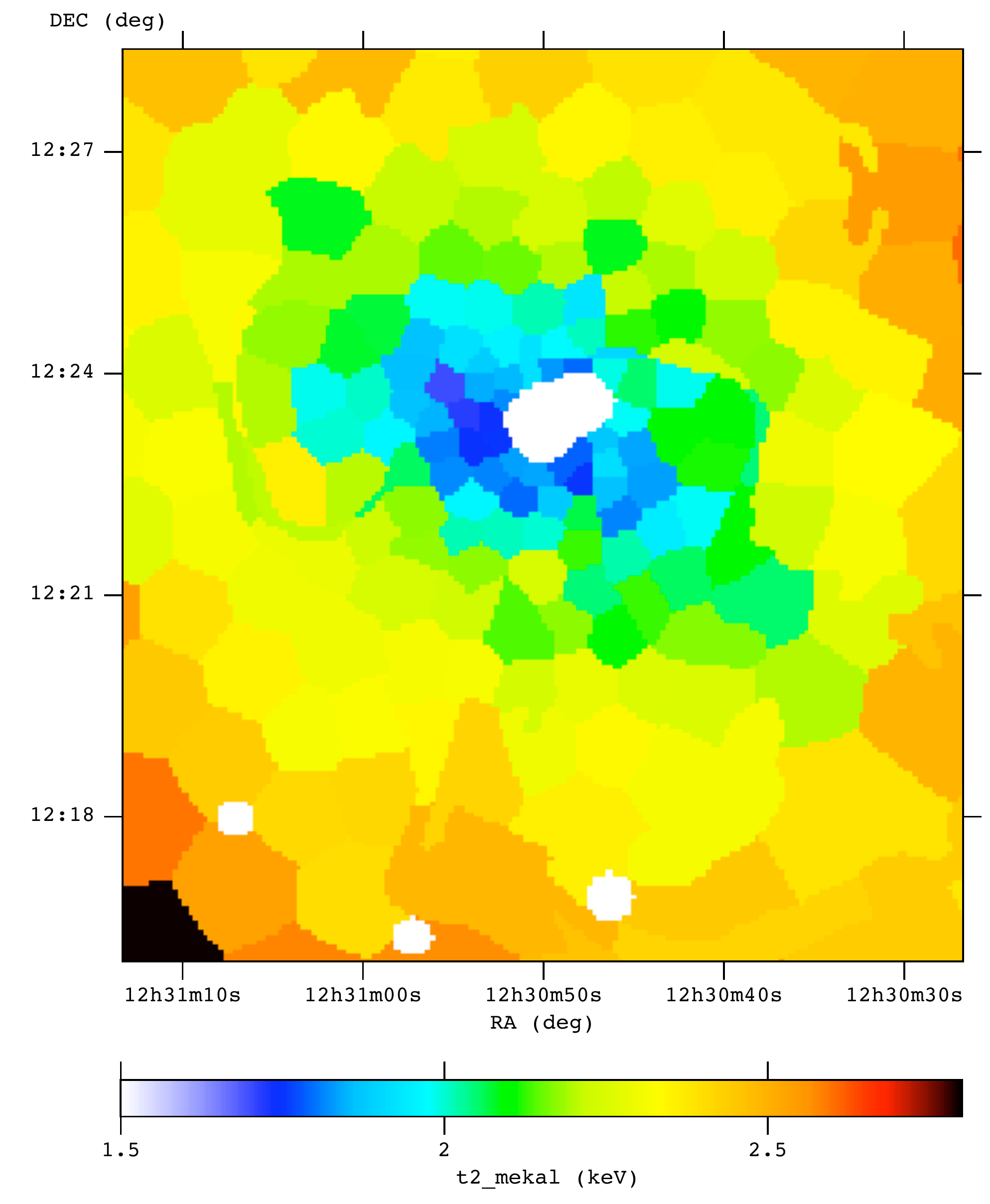}\\
\caption{Temperature of the hot component as determined from the two-temperature MEKAL fit. The map exhibits a fairly regular radial gradient.}
\label{t2_spex}
\end{center}
\end{figure}

Two-temperature models are the simplest next step beyond the single-temperature model approximation. Based on the first XMM-Newton observation of M87, \citet{Belsole01}, \cite{Molendi02} and \cite{Matsushita02} have already shown that the X-ray bright E and SW arms associated with the inner radio lobes are significantly better described by two-temperature models with a cool temperature component at around 1 keV and a hotter component at around 2--2.5 keV. With the new data we are able to analyze in much greater spatial detail the properties of the cool component. 

We find, in agreement with previous work \citep{Molendi02, Matsushita02}, that the temperature of the cool component remains quite stable between 1 and 1.2 keV and shows little or no trends with radius or with surface brightness (Fig. \ref{t1_spex}). For example, the correlation coefficient between the temperature of the cool component and radius in the E arm is only 0.198. The temperature of the hot phase meanwhile exhibits a fairly regular radial gradient (Fig. \ref{t2_spex}). 

\begin{figure}[tbp]
\begin{center}
\includegraphics[width=\columnwidth, bb=18 144 592 718]{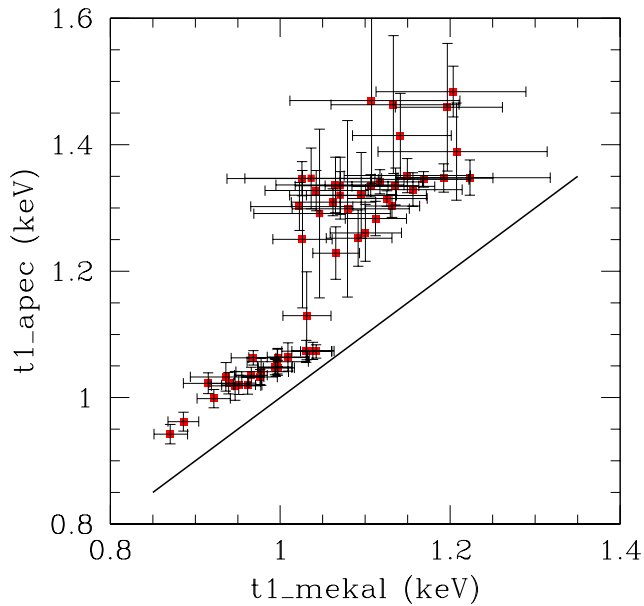}\\
\caption{Differences in determination of the temperature of the cool component with the APEC (fitted in XSPEC) and the MEKAL (fitted in SPEX) models. Errorbars are at the 1$\sigma$ level. The equality line $T_{1,MEKAL}=T_{1,APEC}$ is overplotted.}
\label{mekal_apec}
\end{center}
\end{figure}

For comparison we also performed a similar two-temperature model fit using the APEC model in XSPEC. The differences in determining the temperature of the cool component can be seen in Fig \ref{mekal_apec}. In general, we find very similar trends from both models, however the APEC model seems to give overall higher values than the MEKAL model. This effect is more pronounced for values of $T_{\mathrm{cool,APEC}} > 1.2$ keV than for  $T_{\mathrm{cool,APEC}} < 1.2$ keV. We note that the MEKAL model implemented in SPEX includes a more recent list of transitions than the MEKAL implemented in XSPEC (J. Kaastra, private communication) and that APEC is not implemented in SPEX, thus a direct comparison of the two models is not possible.

\begin{figure}[tbp]
\begin{center}
\includegraphics[width=\columnwidth, bb=35 116 577 677]{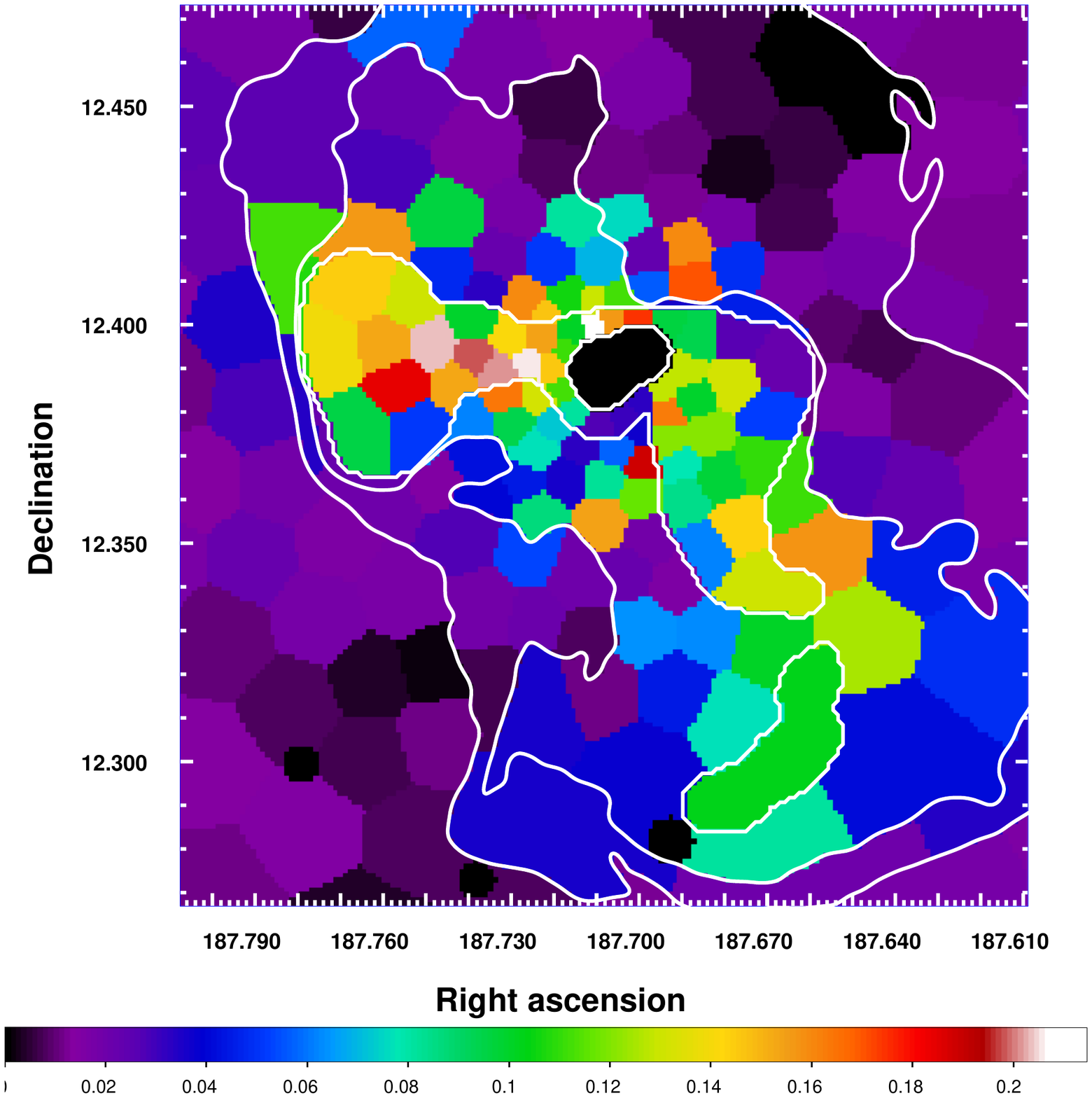}\\
\caption{Map of the fraction of the emission measure of the cool component to the total emission measure (hot + cool component) for the two-temperature MEKAL fit. 90 cm radio contours are overplotted in white.}
\label{n1fract2t}
\end{center}
\vspace{-1cm}
\end{figure}

While the temperature of the cool phase does not seem to vary strongly in the fits, the normalization of the cool gas shows spatial variations correlated with the inner radio lobes. In Fig. \ref{n1fract2t} we plot the fraction of the emission measure of the cool phase to the total thermal emission, $Y_{\mathrm{cool}}/(Y_{\mathrm{hot}}+Y_{\mathrm{cool}})$, overlaid with the 90 cm radio contours \citep{OEK00}. As already noted by e.g. \cite{Belsole01}, the spatial correlation is more pronounced in the E arm, while in the SW arm it seems that the radio plasma is twisting around the uplifted arm.

\subsection{The differential emission measure model}

We also fitted a model with a continuous temperature distribution where the emission measure is a power-law function of the temperature, as described in Sect. \ref{sec:model}. As the model by \cite{Churazov01} also indicates, a multiphase structure with a continuous distribution of the emission measure as a function of temperature is more realistic because, during the rise of the radio lobes, gas at several radii gets entrained and mixed \citep[see also][]{Brueggen02}. The power-law shape of the emission measure as a function of temperature is the simplest model that entails a continuous temperature distribution. 

\begin{figure}[tbp]
\begin{center}
\includegraphics[width=\columnwidth, bb=35 106 577 687]{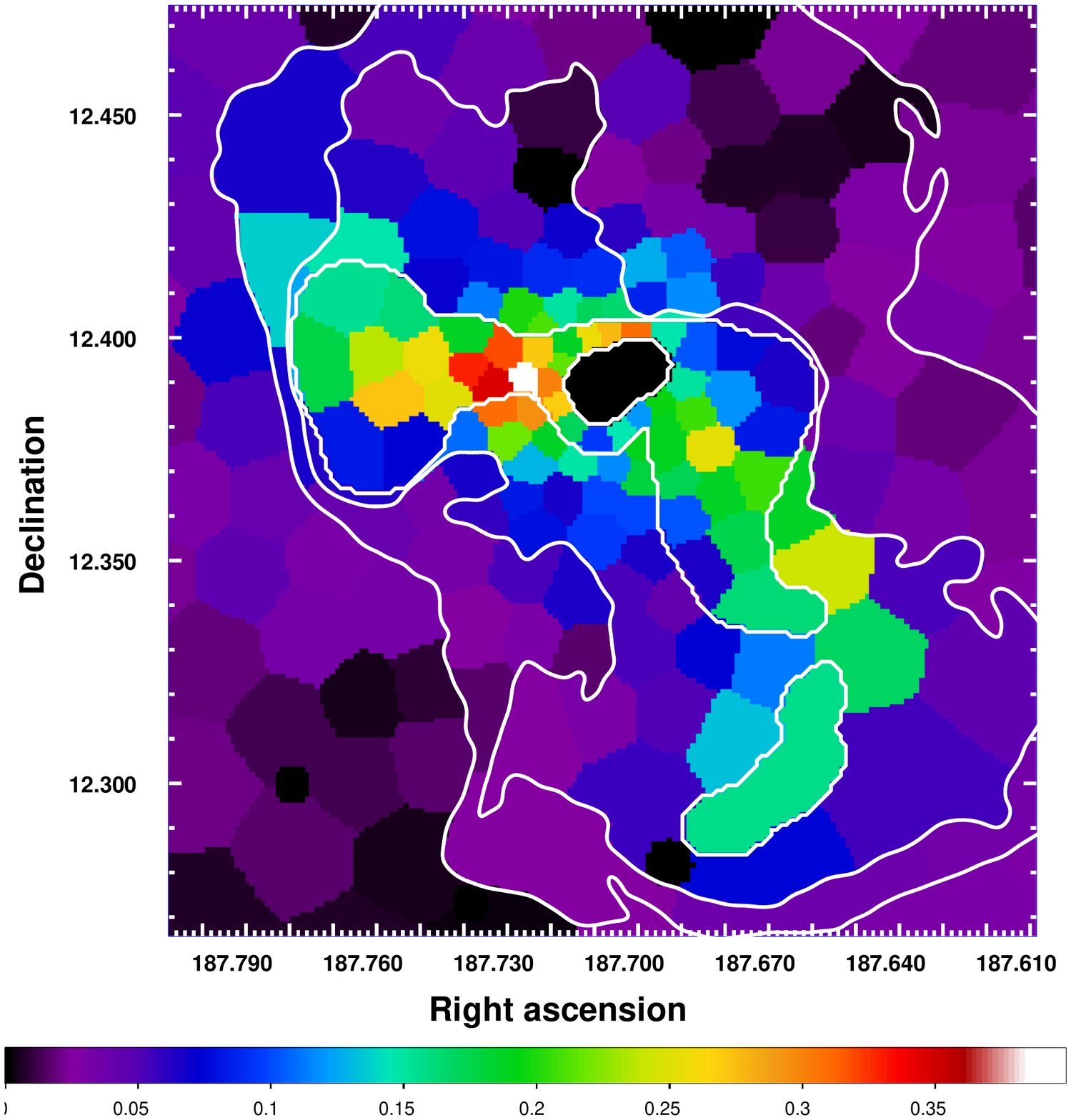}\\
\caption{Fraction of the emission measure of the gas below 1.5 keV to the total emission measure from the {\it{wdem}} fit. 90 cm radio contours are overplotted in white.}
\label{percool}
\end{center}
\end{figure}

\begin{figure}[tbp]
\begin{center}
\includegraphics[width=\columnwidth, bb=18 49 595 743]{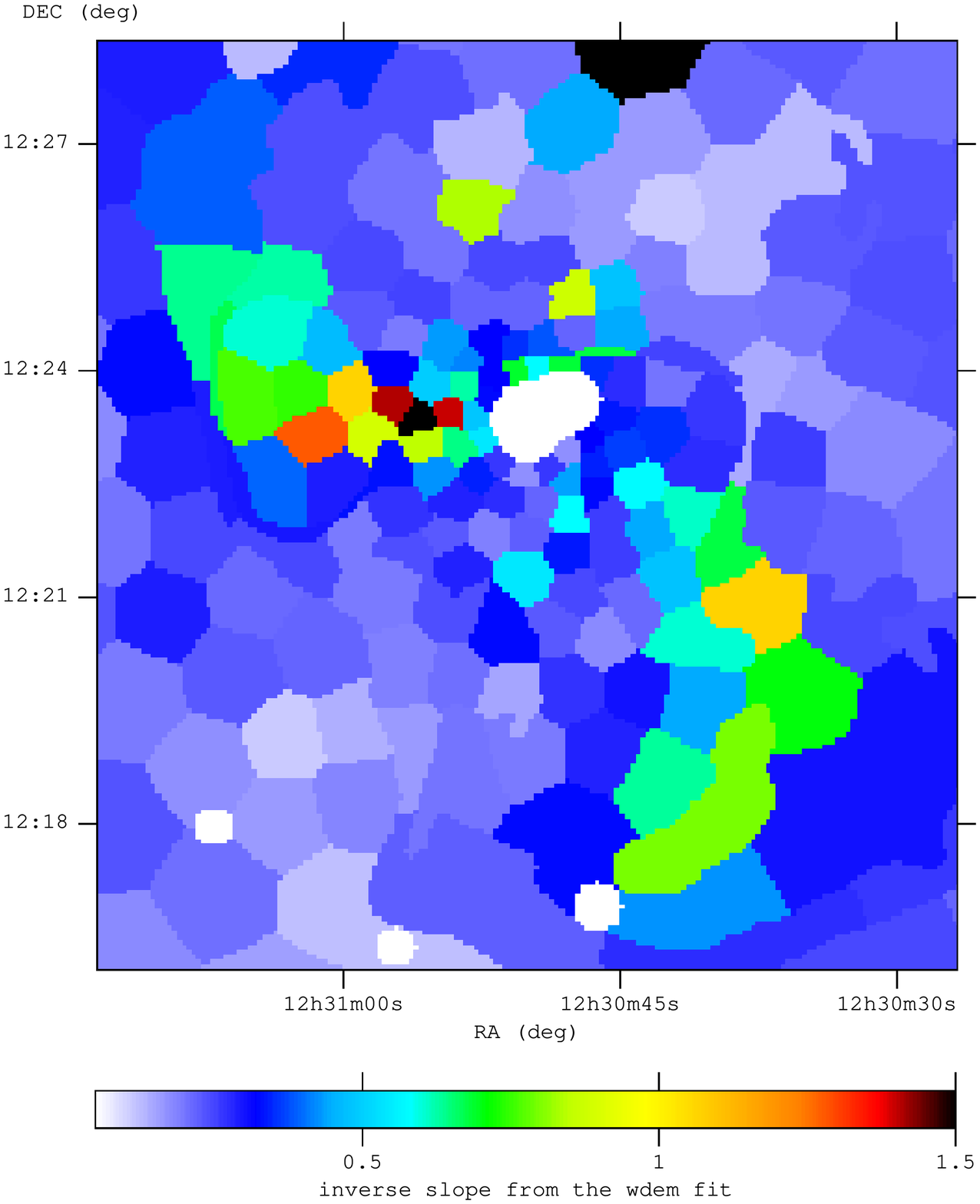}\\
\caption{Map of the inverse slope $\alpha$ from the {\it{wdem}} fit. Higher values in the map reflect regions where the emission measure distribution is flatter.}
\label{slopemap}
\includegraphics[width=\columnwidth, bb=21 50 592 742]{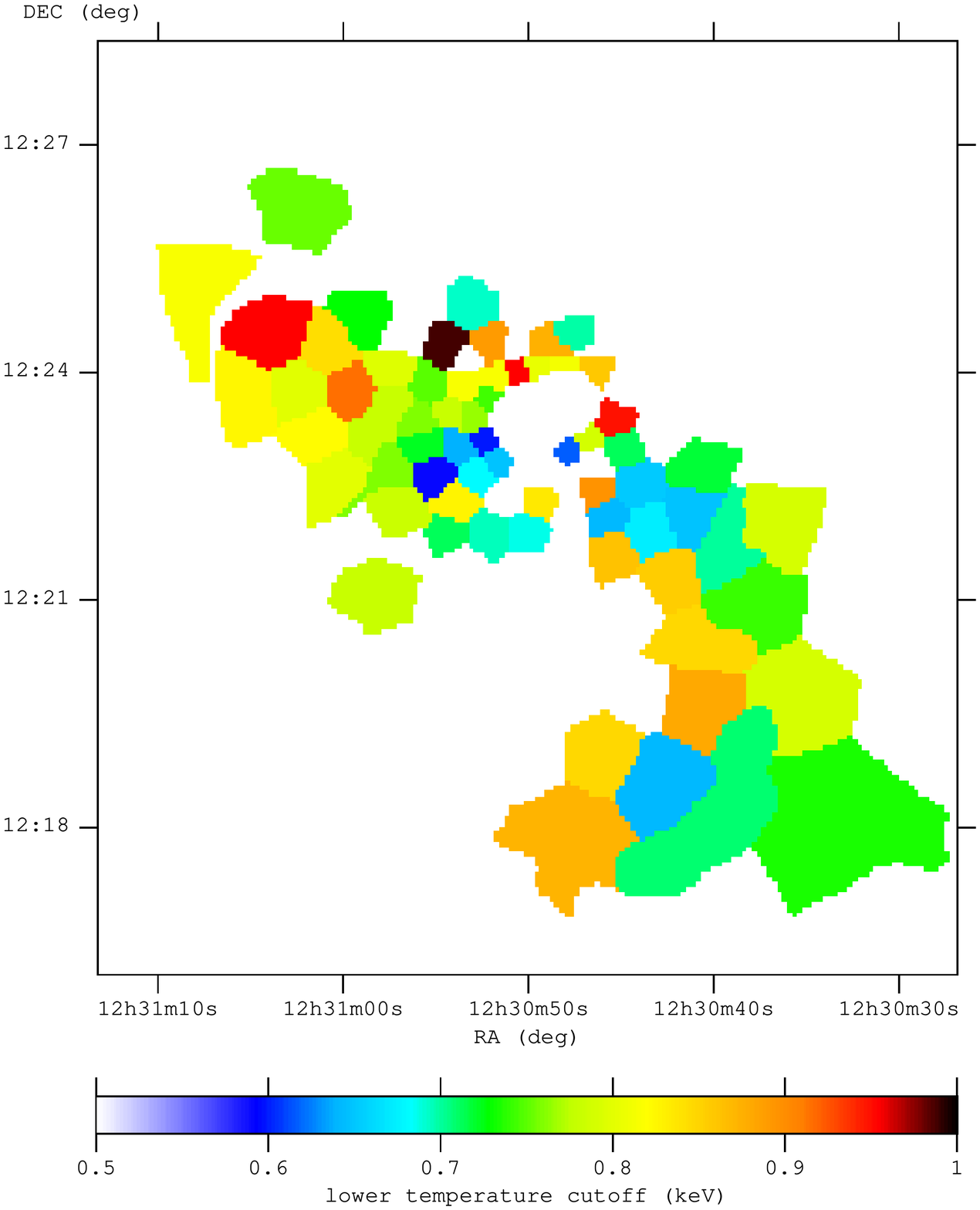}\\
\caption{Map of the lower temperature cutoff from the {\it{wdem}} fit. Emission from gas at as little as 0.6 keV is detected. Only the bins are shown where the fraction of gas below 1.5 keV was more than 3$\sigma$ significant.}
\label{cmap}
\end{center}
\end{figure}

Our main direct result from the {\it{wdem}} fit is shown in Fig. \ref{slopemap}, where we plot the map of the inverse slope $\alpha$ of the power-law emission measure distribution. Higher values in the map indicate regions where the emission measure distribution is flatter, hence where one expects proportionally more cool gas to have been mixed together with the ambient gas. To quantify this more exactly, we calculate from the fit results the fraction of the emission measure of the cool gas to the total emission measure according to Eq. \ref{eq:ycool}. We chose $T_{\mathrm{maxcool}}$ as 1.5 keV, which is roughly the smallest deprojected value of the ambient gas computed by \cite{Matsushita02}. Fig. \ref{percool} shows the spatial distribution of the cool gas fraction, revealing very similar spatial trends as in Fig. \ref{n1fract2t} where the fraction of cool gas in the 2T model was plotted. Thus, the relative contribution of the cool gas to the total emission is similar in both models.

However, the temperature structure of the cool gas can be very different, since the same relative contribution can be achieved with a shallower slope (large $\alpha$) and larger low-temperature cutoff or with a steeper slope but a smaller $cT_{\mathrm{max}}$. In Fig. \ref{cmap}, we plot the map of the lower temperature cutoff $cT_{\mathrm{max}}$ in the regions where we find a more than $3\sigma$ significant fraction of cool gas (in the other regions the gas is not significantly non-isothermal, thus it is irrelevant to speak of a lower temperature cutoff). We find lower cutoff temperatures which go down to as little as 0.6 keV, thus the {\it{wdem}} model allows for the presence of gas much below the best-fit value of the cool temperature in the 2T fit. 

From the fit we find that, based on the chi-square ($\chi^2$) values, we cannot significantly differentiate between the {\it{wdem}} and the two-temperature model. The {\it{wdem}} model that we fitted had the same number of free parameters (upper temperature cutoff, lower temperature cutoff, power-law slope, normalization and element abundances) as the two-temperature fit (two temperatures, two normalizations and element abundances) and the difference between the $\chi^2$ values is between -20 and 10 points for a typical number of degrees of freedom per bin of 1050--1100. 

\subsection{Discussion of the temperature structure and physical feasibility of the different models}

Although based on the improvement of the $\chi^2$ values we cannot determine which model provides a better description of the data, there are several arguments in favor of the continuous temperature distribution.

\begin{figure}[tbp]
\begin{center}
\includegraphics[width=\columnwidth, bb=17 144 560 680]{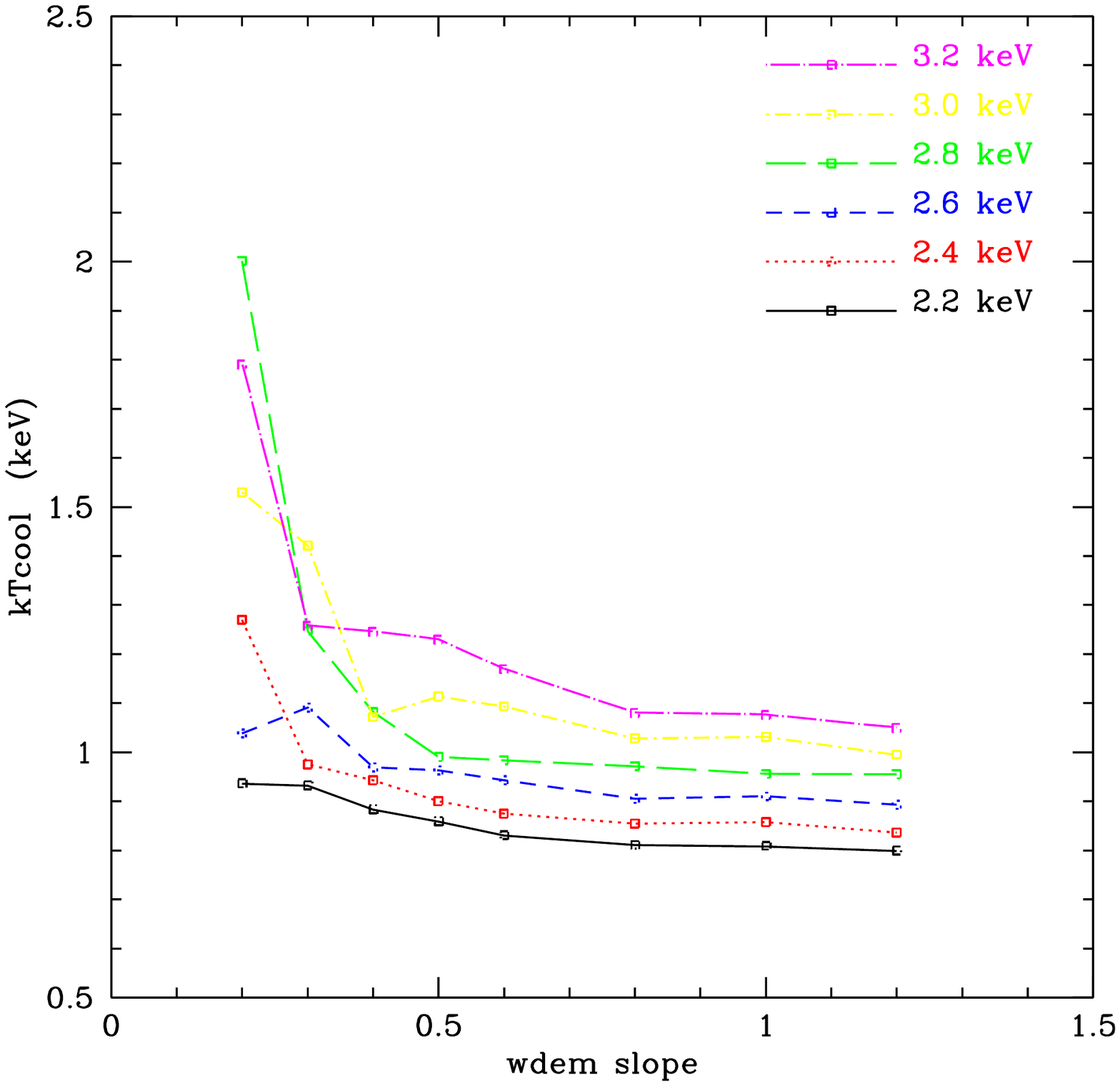}\\
\caption{Results of fitting various simulated {\it{wdem}} spectra with a two-temperature model. The best fit cool temperature in the 2T model remains constant at around 1 keV for a wide range of input parameters in the simulated spectra. The different curves are for different values of the upper temperature cutoff, $T_{\mathrm{max}}$.}
\label{c025}
\end{center}
\end{figure}

Firstly, during the rise of the radio lobes, gas at several radii should get entrained, mixed, and the uplifted gas should expand and cool adiabatically as it moves into regions of lower pressure. Alternatively, the uplifted gas could be heated up by e.g. heat conduction from the surrounding hot phase or by energy dissipation from turbulent motions. In any case, it is very unlikely that the heating and cooling should exactly balance each other to yield the spatially constant temperature of the cool component that we observe in the two-temperature fits (both MEKAL and APEC). More probably, the constancy of the cool temperature could be an artifact of approximating a more complex temperature structure with a two-temperature model. In order to test this, we simulated a series of {\it{wdem}} spectra with different slopes, different high-temperature cutoffs, and a lower temperature cutoff of $0.25 \times T_{\mathrm{max}}$, which is a typical value we find in our fits, and fitted these simulated spectra with a two-temperature MEKAL model. We find indeed, as plotted in Fig. \ref{c025}, that the best fit temperature of the cool component in the 2T fit is constant at around 1 keV for a wide range of slopes and high-temperature cutoffs. Thus the fact that the cool temperature in the 2T fit is so constant is very likely an indication that the 2T fit is only an approximation of a more complex continuous temperature distribution. We note that very similar results are also found for lower temperature cutoffs of between $\sim$ 0.2--0.5 $\times T_{\mathrm{max}}$.

Secondly, \cite{Werner06} detect \ion{Fe}{xvii} lines in the RGS spectra of M87, lines which are emitted between ~0.14 and ~0.85 keV \citep{Arnaud92}. The presence of these lines is inconsistent with the results of the two-temperature fits (both MEKAL and APEC) and can only be accounted for with a multi-temperature model which predicts emission from gas with temperatures of as low as 0.6 keV. Moreover, the RGS results show the presence of gas at $\sim 0.7$ keV out to $3.5^\prime$ in the SW arm, which also speaks in favor of the {\it{wdem}} model.

Our data thus points towards the conclusion that a continuous differential emission measure model for the gas in the X-ray arms is likely to be a more physical and more realistic description than a two-temperature plasma approximation. In most of the following sections we will thus primarily focus on the results of the {\it{wdem}} fit.

\subsection{Discussion of the implications of multi-temperature structure for thermal conduction}\label{sect:spitzer}

A gas parcel with volume $V$ and number density $n$, in thermal contact over an area $A$ with an ambient gas with a temperature which is different by $\Delta T$, will reach thermal equilibrium with this ambient gas through heat conduction on a time scale that can be estimated as \citep[adapted from][]{Boehringer89}:

\begin{equation}\label{tau_hc}
\tau_{hc}\approx\frac{\Delta H}{q A}\approx\frac{\frac{5}{2}nk_B\Delta T}{f \kappa_0 T^{5/2} \nabla T}\frac{V}{A}
\end{equation}
where $\Delta H$ is the enthalpy which must be transfered to the gas parcel so that it reaches thermal equilibrium with the ambient gas, $q$ is the heat flux according to \citet{Spitzer62}, $\kappa_0$ is the Spitzer heat conduction coefficient, $\kappa_0 \sim 6 \times 10^{-7} \rm{erg \:cm^{-1}\: s^{-1}\: K^{-7/2}}$, and $f$ is a coefficient allowing for the reduction in the heat conduction due to the effects of the magnetic field. For the case of a continuous multi-temperature distribution as the one described by the {\it{wdem}} model, the calculation of the heat conduction flux and hence the heating time scale is generally more complex. We can however make simplifying assumptions which will lead to upper limits on the allowed heat conduction such that the cool gas detected survives for over $\sim 10^7$ years, which is the estimated age of the inner radio-lobes  \citep[e.g.][]{Forman06}. The temperature gradient, and thereby also the heat flux, is smallest if we assume that the gas is stratified in each spatial bin such that gas at the coolest temperatures is furthest away from gas at the hottest temperatures. Moreover, $V/A$ is largest assuming a spherical geometry. Combining these two assumptions and the quoted age of the radio lobes we arrive at an upper limit for the allowed $f$.

We aim to rewrite Eqn \ref{tau_hc}, which holds for two gas phases in thermal contact, for the case of the multi-temperature distribution described by the {\it{wdem}} model. For this we will calculate how much enthalpy $\Delta H$ must be transferred to the gas contained in the {\it{wdem}} model below a chosen $T_*$ in order to heat this gas up to $T_*$. This is given by:

\begin{equation}\label{delH}
\Delta H=\frac{5}{2}\int^{T_*}_{\rm{cT_{max}}}(n_e+n_i)k_{\rm B}(T_*-T)\:dV .
\end{equation}
Note that according to the continuous distribution of the {\it{wdem}} model, $T_*$ can be arbitrarily close to $cT_{\rm max}$, although it is probably not physically feasible for the separation to be infinitesimal. Given that $dY = (dY/dT) dT = n_en_i dV$ and assuming pressure equilibrium in the form $(n_e+n_i)\: T=\rm{const}$, it follows that 
\begin{equation}\label{dvdt}
dV \propto T^{2+1/\alpha} dT.
\end{equation}
The proportionality constants can be calculated assuming a total volume for each bin (see Sect. \ref{sect:mass}). The volume $V$ in Eqn \ref{tau_hc} can then be calculated integrating $dV$ from $cT_{\rm max}$ to $T_*$. Assuming the gas below $T_*$ to be inside a sphere, one can then calculate the area $A$ corresponding to this $V$. Finally, $\nabla T=dT/dr=dT/dV \cdot dV/dr$ can be evaluated at $T_*$ using the assumption of spherical geometry and Eqn \ref{dvdt}. This allows the calculation of $\tau_{hc}$, and the requirement that $\tau_{hc}(T)<10^7$ years for various $T_*$ can help us place upper limits on $f$. However, we cannot account for possible mechanisms that would change the temperature cutoffs or the slope of the emission measure distribution since the time it was produced. Such mechanisms include mixing and the fact that gas at very low or very high temperatures, which could have been present at the beginning, has already been heated/cooled and is no longer detected.

\begin{figure}[tbp]
\begin{center}
\includegraphics[width=\columnwidth, bb = 18 144 592 718]{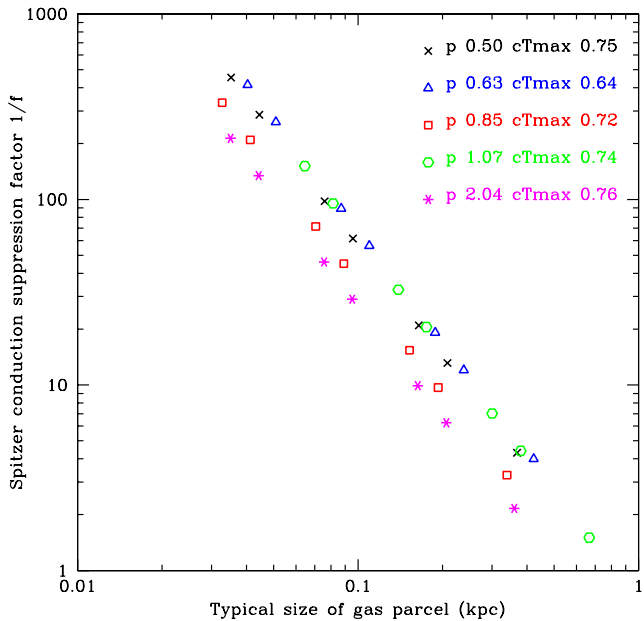}\\
\caption{Factor by which the Spitzer heat conduction must be suppressed (this corresponds to $1/f$ in Eqn. \ref{tau_hc}) in order to maintain the cool gas in the multi-temperature structure over at least $10^7$ years, as a function of the size corresponding to the volume of the cool gas, $V(<T_*)$. The {\it{wdem}} inverse slope denoted here by $p$ and lower temperature cutoff $cT_{\rm max}$ (in keV) of the 5 different bins used for the calculation are indicated in the legend.}
\label{spitzer}
\end{center}
\vspace{-0.5cm}
\end{figure}

We plot in Fig. \ref{spitzer} the conductivity suppression factor $1/f$ needed to maintain the cool gas in the multi-temperature structure over at least $10^7$ years against the radius $r \sim (3V/4\pi)^{1/3}$ corresponding to $V(<T_*)$. For the calculation we used $T_*-cT_{\rm max}$ in the range from $5\times10^2$--$5\times10^5$ K. We calculated the values plotted in the figure using results from 5 different bins in the E and SW arms spanning values of $\alpha$ between 0.5 and 2. All our bins had similar low-temperature cutoffs of $cT_{\rm max} \sim$0.64--0.76 keV. We find that, for typical blob sizes of 0.1 kpc, similar to the widths of the cool filaments seen in the Chandra image \citep[the filaments do not appear in images produced in energy bands above 2 keV,][]{Forman06}, a suppression factor on the order of $\sim$30--100 is required.

%________________________________________________________________

\section{The spatial distribution of metals}\label{sect:spatial_distr_metals}
%_________________________________________________________________

\subsection{Abundance maps}\label{sect:abundmaps}

The spectra in each spatial bin provide enough statistics not only to test the multi-temperature structure of the gas but also to determine the abundances of several heavy elements and to analyze their spatial distribution and the influence of different multi-temperature models on the abundance determination. In the spectral fits we left the abundances of O, Mg, Si, S, Ar, Ca, Fe and Ni free. The Ne abundance was fixed to the Mg abundance, since the Ne line falls within the Fe-L line complex and thus the Ne abundance cannot be accurately determined with the spectral resolution of the EPIC detectors.

\begin{figure}[tbp]
\begin{center}
\includegraphics[width=\columnwidth, bb=36 117 577 675]{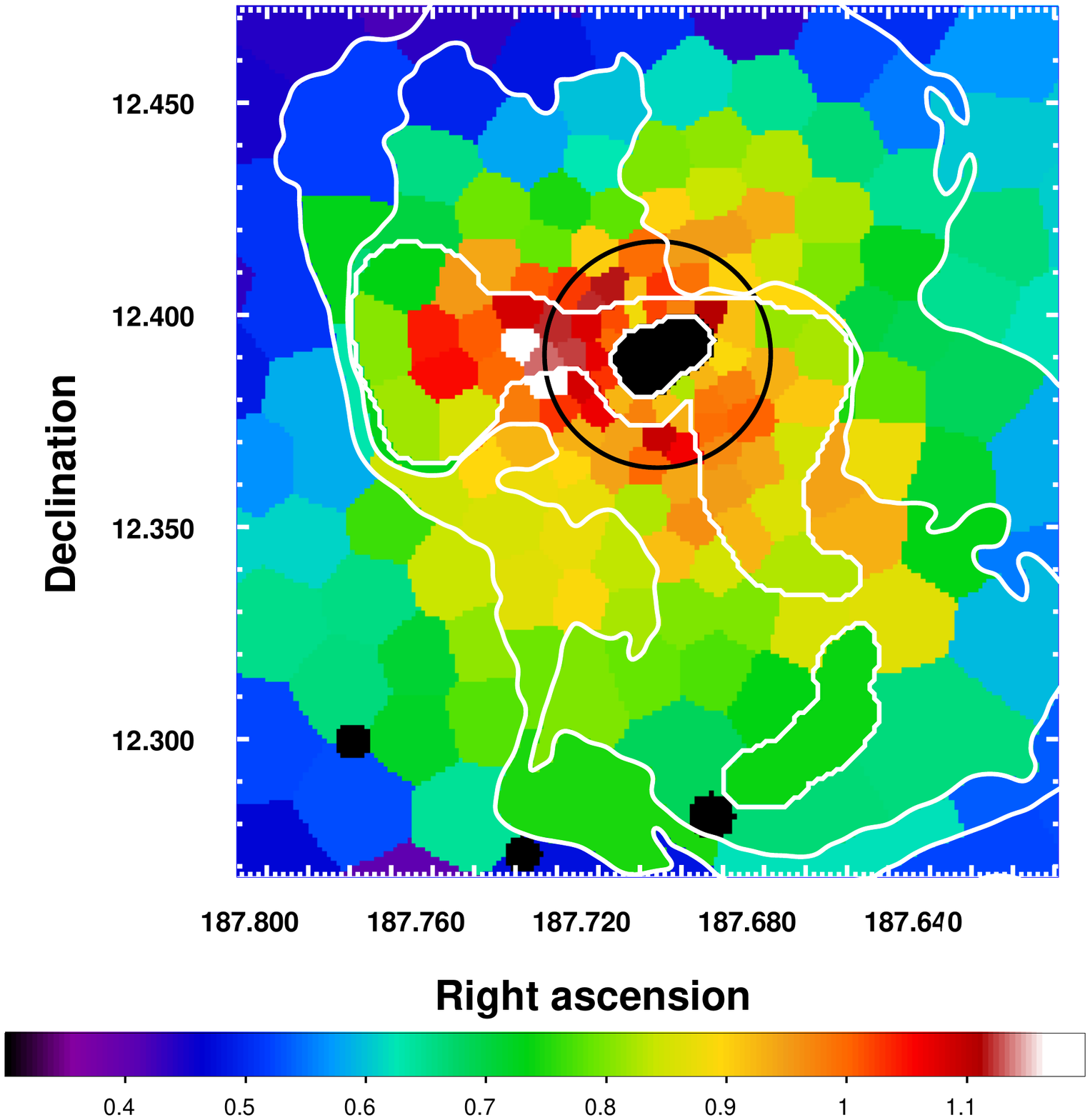}\\
\caption{Fe map determined from the {\it{wdem}} model. Beyond the expected radial gradient, one clearly sees the enhancement in Fe abundance in the arms, especially within the E radio lobe. The half-light radius of M87 is marked with a black circle, the 90 cm radio contours are over-plotted in white.}
\label{femapwdem}
\end{center}
\vspace{-0.5cm}
\end{figure}

Because of the good statistics around the Fe-L line complex, the Fe abundance is usually determined with the lowest statistical errors. We present a map of the Fe abundance determined from the {\it{wdem}} model in Fig. \ref{femapwdem}. As expected from previous work \citep[e.g.][]{Matsushita03}, a radial gradient is apparent. Beyond this radial trend however, deviations associated with the inner radio lobes can be seen as clear evidence of the influence of the AGN on the spatial distribution and transport of metals. The best region to illustrate this is the eastern lobe, which from the ``mushroom-cloud'' shape in the radio image appears as a more undisturbed system than the SW lobe. Here, one clearly sees a stem-shaped enhancement in the Fe abundance which coincides very well with the rising plasma bubble. The SW lobe also presents a higher metallicity with respect to the average radial profile, but this enhancement seems a bit less concentrated than in the E lobe and is more difficult to see from the map in Fig. \ref{femapwdem}. A more detailed quantitative analysis will be presented in Sect. \ref{sect:fepercor}.

\begin{figure}[tbp]
\begin{center}
\includegraphics[width=\columnwidth, bb=18 144 592 718]{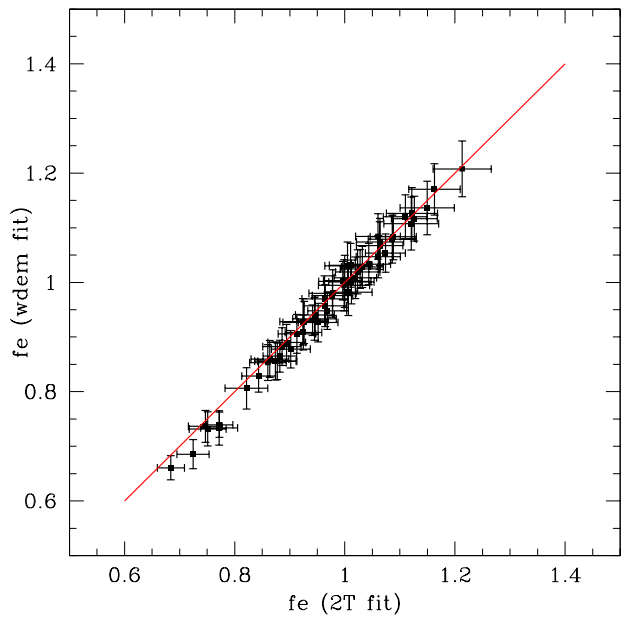}\\
\caption{Fe abundance determination with two different multi-temperature models. The red line represents the function Fe({\it{wdem}})=Fe(2T). It is easily seen that the abundance determinations from the two models agree very well, with a very low scatter. For clarity, only the points are plotted where the emission measure of the cool component in the 2T fit and the fraction of cool gas in the {\it{wdem}} fit were more than $3\sigma$ significant. We note, however, that the correlation is equally strong also for all the other points in the maps.}
\label{fewdem2t}
\end{center}
\end{figure}

Since the Fe-L complex is very sensitive to the temperature of the gas, one might question if and how the determined Fe abundance systematically depends on the model involved. To answer this, we plot in Fig. \ref{fewdem2t} the Fe abundance determined from the 2T and the {\it{wdem}} models. Reassuringly, the correlation between the two sets of fit values is very tight, which allows us to be confident in the measured abundance trends.
\subsection{Metal abundance patterns of different elements in and outside the multi-temperature regions}\label{sect:metalpatterns}

The spatial distributions of other elements, especially S and Si which have the next best determined abundances, are very similar to that of Fe. Thus, we chose not to include 2D maps for these elements but rather to plot their abundance trends relative to Fe. We find that, while the overall abundance in the arms is higher, the Fe, Si, S and O abundances correlate well both inside and outside the radio lobe regions. The scatter in the O/Fe relation is larger than for Si/Fe and the O/Fe slope is much shallower, but we do see a small increase in the O abundance for higher Fe values. Therefore the cool gas must be enriched also with oxygen, for example through stellar mass loss in the galaxy. 
\begin{figure}[tbp]
\begin{center}
\includegraphics[width=\columnwidth, bb=18 144 592 718]{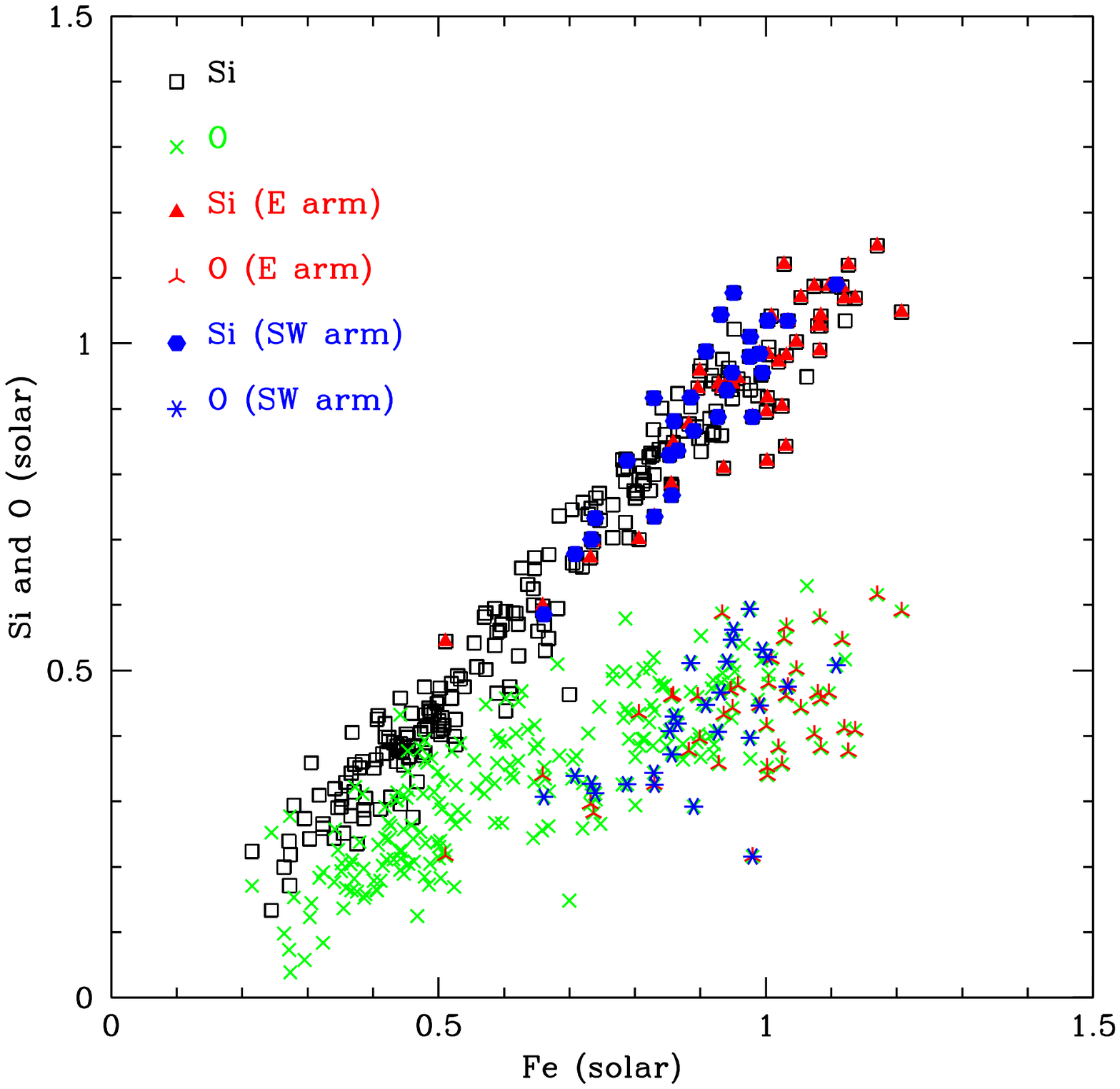}\\
\caption{Oxygen and silicon vs. iron abundances in and outside the arm regions ({\it{wdem}} fit). Both the Si vs. Fe and O vs. Fe in the arms (marked with special symbols, see graph legend) follow the same trends, respectively, as the Si vs. Fe and O vs. Fe data points for bins outside the arms. Thus, the same abundance patterns are seen in all environments.}
\label{patterns}
\end{center}
\end{figure}
Fig. \ref{patterns} shows a plot of the Si and O abundances against Fe, both for the bins where the cool gas fraction was significant (marked with special symbols) and for the rest of the bins in the fit. Since the Si and S abundances are very similar, we do not add the S/Fe data points to the plot for legibility. Both for Si/Fe and O/Fe we see that the trend in the multi-temperature regions coincides well with the trend outside the influence of gas-uplift by the radio lobes \citep[note that][also find a constant ratio of O/Fe with radius in the inner $9^\prime$ of M87]{Gastaldello02}. To quantify this, we fitted a line to the O vs. Fe, Si vs. Fe and S vs. Fe data points in and outside the arm regions. The best fit slopes are given in Table \ref{abundslopes}. These slopes are expected to become different at low values of the Fe abundance, where the contribution by SNeIa goes to zero and the O/Fe and Si/Fe ratios are determined by the pre-enrichment patterns of SNeII \citep[for detailed discussion of this see, e.g.,][]{Matsushita03}. With the bins included in our plot in Fig. \ref{patterns} however, we do not seem to reach down to low enough Fe abundances to see this effect significantly.

\begin{table}[htdp]
\caption{Comparison of best-fit metal abundance ratios in and outside the multi-temperature regions ({\it{wdem}} fit).}
\begin{center}
\begin{tabular}{|c|c|c|}
\hline
 &inside & outside \\
 &multi-temperature region &multi-temperature region\\
 \hline
 O/Fe & 0.44 $\pm$ 0.07 & 0.49 $\pm$ 0.03\\
\hline
 Si/Fe& 1.02 $\pm$ 0.06& 1.10 $\pm$ 0.02\\
 \hline
 S/Fe & 1.05$\pm$ 0.09 & 1.18 $\pm$ 0.03\\
 \hline
\end{tabular}
\end{center}
\label{abundslopes}
\end{table}
As seen in Table \ref{abundslopes}, there are indications that Fe is slightly more abundant in the multi-temperature regions (O/Fe, Si/Fe and S/Fe are all smaller than in the single-temperature regions), which would imply a more important relative contribution from SNeIa, but none of these differences is more than 2$\sigma$ significant. Because of this and since the O vs. Fe and Si vs. Fe data points in the multi-temperature regions overlap well with the trend outside the influence of gas-uplift by the radio lobes, we conclude that the AGN, at least at this advanced stage of the evolution of the galaxy, enriches the gas halo without altering the relative abundance patterns.

%________________________________________________________________

\section{The properties of the cool gas}\label{sect:coolgas}
%________________________________________________________________

\subsection{The correlation between the amount of uplifted cool gas and metallicity}\label{sect:fepercor}

Several authors have discussed the AGN/ICM interaction as a driver for metal-transport from the central galaxy into the ICM \citep[for examples of recent work, see e.g.][]{Rebusco06, Roediger06}. Comparing Fig. \ref{femapwdem} to \ref{n1fract2t} or \ref{percool}, a possible mechanism by which the AGN influences the distribution of the metals in the gas halo is readily apparent: if the cool gas from the center of the galaxy is richer in metals, then the uplift of this gas by the radio bubbles will also enable the transport of the metals towards the outskirts of the cluster. Indeed, it seems that regions of enhanced metallicity in the lobes correspond well to regions where the fraction of the cool component is large. 

\begin{figure}[tbp]
\begin{center}
\includegraphics[width=\columnwidth, bb=18 144 592 718]{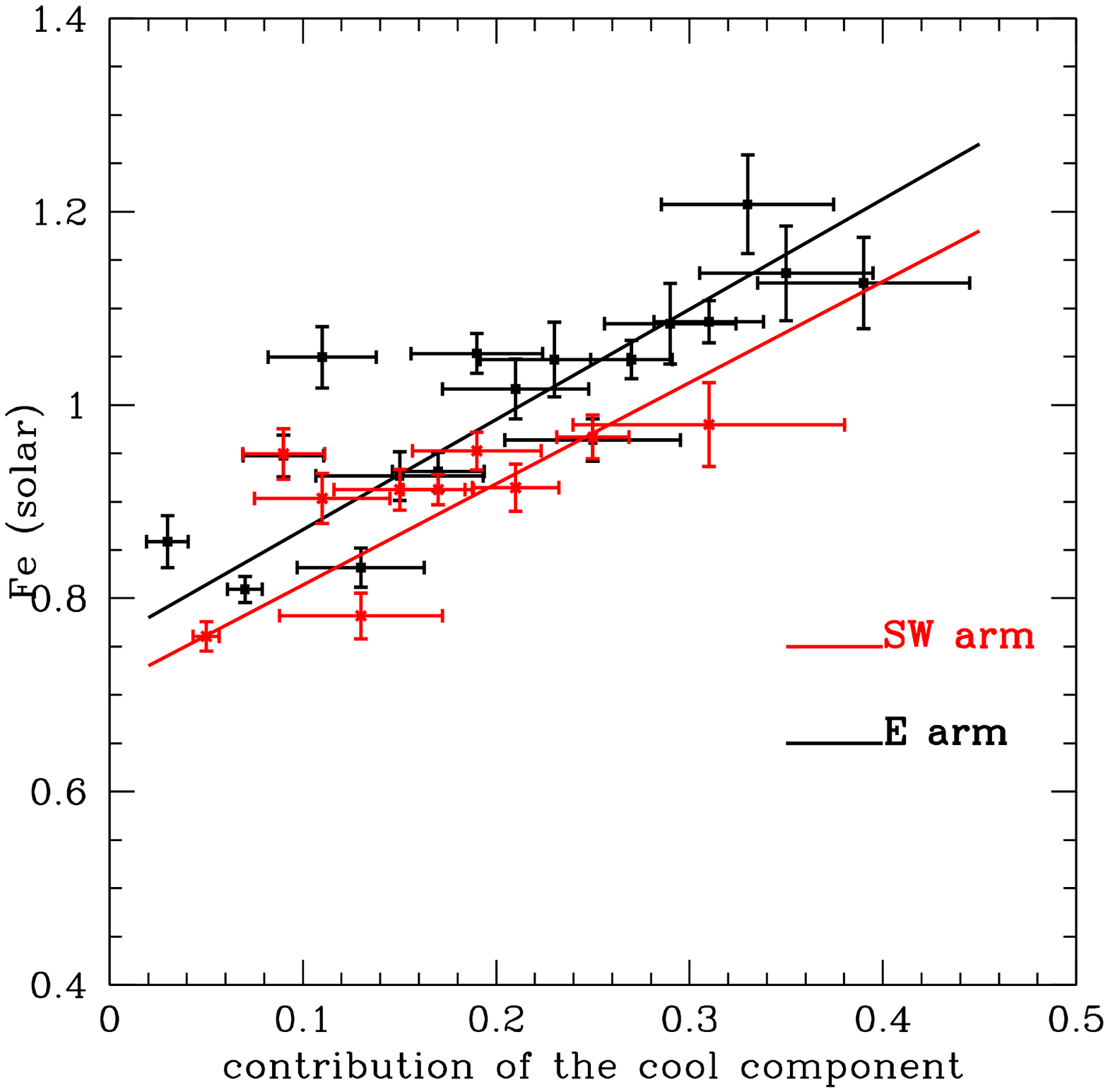}\\
\caption{Averaged Fe abundance in 2\%-wide bins of the fraction of the cool component and linear fit for each arm. The linear correlation coefficients are 0.86 for each of the two arms, indicating a very clear correlation.}
\label{percool_vs_fe}
\end{center}
\vspace{-0.75cm}
\end{figure}

To illustrate this correlation, we plot in Fig. \ref{percool_vs_fe} the Fe abundance as a function of the fraction of the cool component in the {\it{wdem}} model for the regions where this fraction was more than 3$\sigma$ significant, for the E and SW arm separately. In the plot, the Fe abundance was binned in 2\%-intervals of the cool gas fraction. We calculated the linear correlation coefficient for each arm without binning, taking into account the error bars on both axes \citep{Akritas96}, and obtained a value of 0.66 for the E arm and 0.42 for the SW arm. Combining the data points for both arms we obtain a correlation coefficient of 0.48. This confirms a moderate linear dependence of the measured Fe abundance on the fraction of the cool component overall. Furthermore, binning the Fe abundance in 2\%-intervals of the fraction of the cool component and recalculating the linear correlation coefficients, we obtain a value of 0.86 for each of the two arms, indicating a very clear correlation. The linear fits were performed taking into account the errors on both axes. Extrapolating the best-fit line equations to 100\% cool-component yields an average iron abundance of $1.88\pm0.18$ and $1.93\pm0.12$ solar for the SW and E arm respectively (quoted errors are at the $1\sigma$ level). 

We adapted this procedure also for the results of the two-temperature fits and obtain comparable results for the mean iron abundance of the cool gas: $2.13\pm0.54$ for the SW arm and $2.38\pm0.29$ for the E arm. However, the correlation is not as strong as in the case of the {\it{wdem}} model. The correlation coefficients after binning the data in 2\%-wide bins of the cool gas fraction are only 0.63 for the E arm and 0.34 for the SW arm. The fact that the Fe abundance correlates much better with the fraction of cool gas determined from the {\it{wdem}} model than with the fraction determined from the 2T model may be an additional indication that a continuous temperature distribution is a better description of the temperature structure in the regions associated with the radio lobes. 

\begin{figure}[tbp]
\begin{center}
\includegraphics[width=\columnwidth]{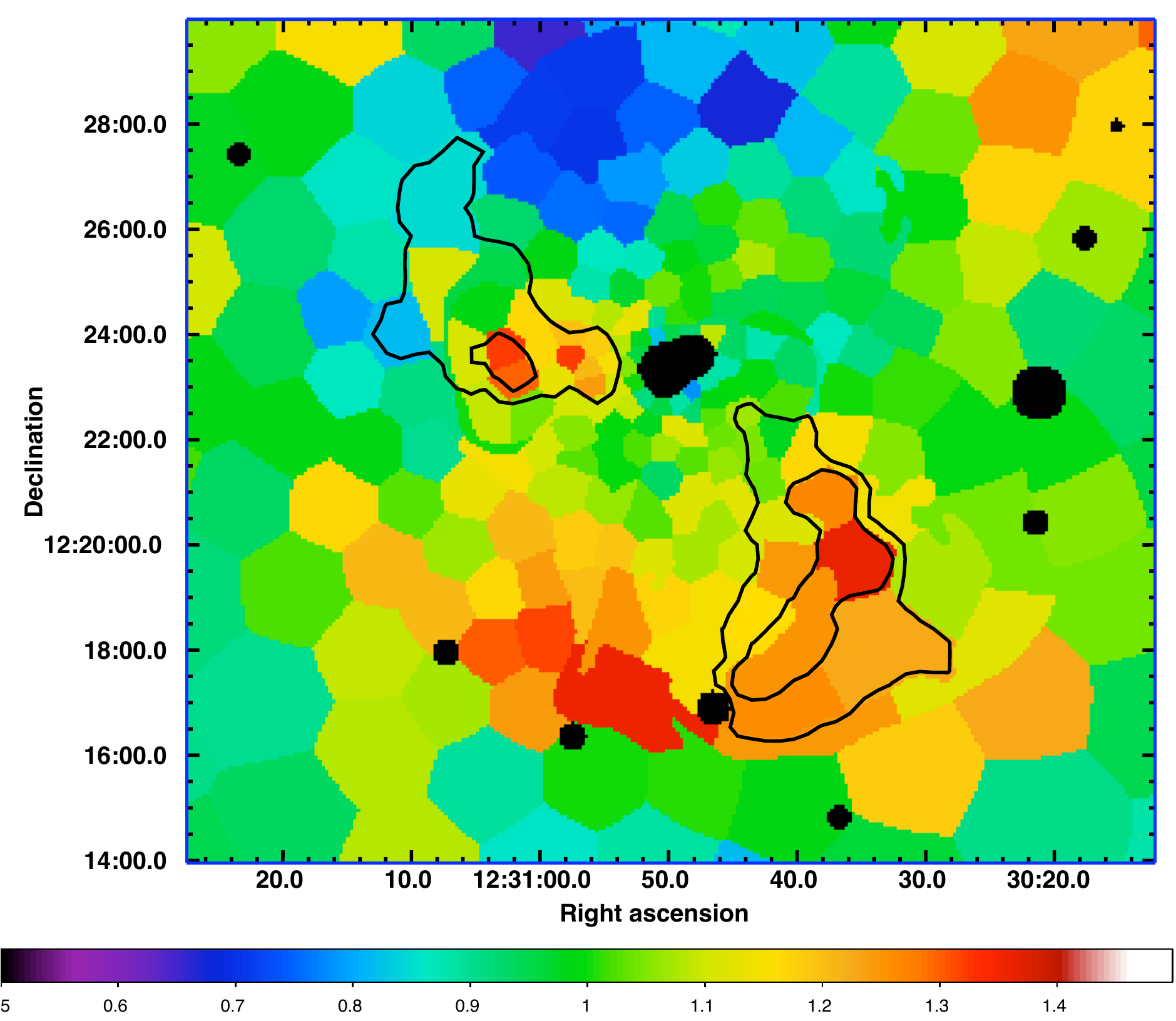}\\
\caption{Map of Fe deviations from a smooth radial profile, $\rm Fe/<Fe>$. Contours of the cool gas fraction deviations from radial symmetry are over-plotted. It is clear that the correlation between Fe and $f_{\rm cool}$ is in most part not due to common radial dependences. The high-abundance feature to the SE corresponds to the cold front presented in \citet{Simionescu07}. }
\label{fedev}
\end{center}
\vspace{-0.75cm}
\end{figure}

To check to which extent the correlation presented in Fig. \ref{percool_vs_fe} may be due to similar radial trends of the cool gas fraction and Fe abundance, we furthermore fitted the radial Fe and $f_{\rm cool}$ profiles from the {\it{wdem}} results using non-parametric, locally weighted, linear regression smoothing, and divided the Fe and $f_{\rm cool}$ maps by the resulting radial models. This should remove any trends due to common radial dependences. We plot in Fig. \ref{fedev} the map of Fe deviations from a smooth radial profile, $\rm Fe/<Fe>$, and over-plot contours of the $f_{\rm cool}$ deviations from radial symmetry, $f_{\rm cool}/<f_{\rm cool}>$. The figure shows a tight correlation between regions of relatively higher iron abundance and relatively higher fraction of cool gas with respect to the radial average. The only exception to this is a high-abundance region to the SE, which does not correspond to a high cool-gas fraction but where \citet{Simionescu07} have already discovered the presence of a cold front associated with a metallicity jump. The relatively low-abundance region opposite the SE feature is probably due to the fact that the higher abundances over a relatively large azimuthal range in the cold front pull up the average at that particular radius, so dividing ``normal'' values at that radius by this average will result in a depression. We calculated the linear correlation coefficients between $\rm Fe/<Fe>$ and $f_{\rm cool}/<f_{\rm cool}>$ and obtain 0.70 for the E and 0.72 for the SW arms (unbinned). The fact that these coefficients are even stronger than for the simple Fe vs $f_{\rm cool}$ correlation indicates that the underlying radial dependence of the abundance of the hot halo may be responsible for a large part of the scatter in Fig. \ref{percool_vs_fe}. We will investigate this further using spectral simulations in the next section.

\subsection{The metallicity of the cool gas}\label{sect:sim}

To explain the correlation between the amount of uplifted cool gas and metallicity presented in the previous section we can begin by assuming that if we fit a multiphase gas for which the Fe abundance is some function of the temperature with a spectral model with a single metallicity, the resulting Fe abundance from the spectral fit can be (neglecting for now e.g. the dependence of the Fe line strength on temperature) approximated by an emission-weighted average:
\begin{eqnarray}
Z_{\rm{Fe}}^\prime&\approx&\frac{\int^{T_{\rm{max}}}_{cT_{\rm{max}}} \rm{Z_{Fe}(T)Y(T)dT}}{\int^{T_{\rm{max}}}_{cT_{\rm{max}}} \rm{Y(T)dT}} \approx \left< Z_{\rm{Fe_{\rm{cool}}}}\right > f_{\rm{cool}}+\left< Z_{\rm{Fe_{\rm{hot}}}}\right > f_{\rm{hot}}\\
&\approx& \left< Z_{\rm{Fe_{\rm{cool}}}}\right > f_{\rm{cool}}^\prime+\left< Z_{\rm{Fe_{\rm{hot}}}}\right > f_{\rm{hot}}^\prime \\
&\approx& \left< Z_{\rm{Fe_{\rm{hot}}}}\right > + \left (\left< Z_{\rm{Fe_{\rm{cool}}}}\right > - \left< Z_{\rm{Fe_{\rm{hot}}}}\right > \right) f_{\rm{cool}}^\prime
\end{eqnarray}
where the quantities denoted by the prime are the results of the fitting while the other quantities are the "true" values characterizing the gas, and it holds that $f_{\rm{cool}}^\prime + f_{\rm{hot}}^\prime = 1$ since $f_{\rm{cool}}^\prime$ and $f_{\rm{hot}}^\prime$ denote the fitted relative emission measure contributions from the cool and hot gas respectively. The cool gas is, as before, defined as gas with a temperature below 1.5 keV, which is the lowest value found in the deprojected temperature profile outside the X-ray arms \citep{Matsushita02}.

This means that the measured Fe abundance $Z_{\rm{Fe}}^\prime$ depends in first-approximation linearly on the fraction of the cool, metal-rich gas $f_{\rm{cool}}^\prime$, or equivalently on $Y_{\mathrm{cool}}/Y$, which is in agreement with what we observe. If this approximation holds, then the Fe abundances obtained by extrapolation to 100\% cool component contribution, as calculated above, are the real values of the Fe abundance of the cool component.

To investigate the validity of this simple approximation, we simulated spectra combining a {\it{wdem}} component between $c\:T_{\mathrm{max}}$ and 1.5 keV with Fe = 2 solar and a second {\it{wdem}} component between 1.5 keV and $T_{\mathrm{max}}$ with Fe = [0.6,0.8,1.0] solar.  The two {\it{wdem}} components had the same slopes and their relative normalizations were chosen such that $dY/dT$ at 1.5 keV was continuous. We stepped the upper temperature cutoff $T_{\mathrm{max}}$ between [2.2,2.6,3.0] keV, the lower temperature cutoff between [0.2,0.25,0.3,0.4]$T_{\mathrm{max}}$ and the slope between 0.2 and 1.2 in intervals of 0.2. This parameter space is equivalent to a range of cool gas contributions from 0 to 60\% of the total emission measure. We assumed an O/Fe ratio of 0.4 and Si/Fe and S/Fe ratios of 1.0 solar (see Table \ref{abundslopes}) for both of the simulated model components. The simulated spectrum was then fitted with a single {\it{wdem}} model with a single abundance.

\begin{figure}[tbp]
\begin{center}
\includegraphics[width=\columnwidth, bb=18 144 592 718]{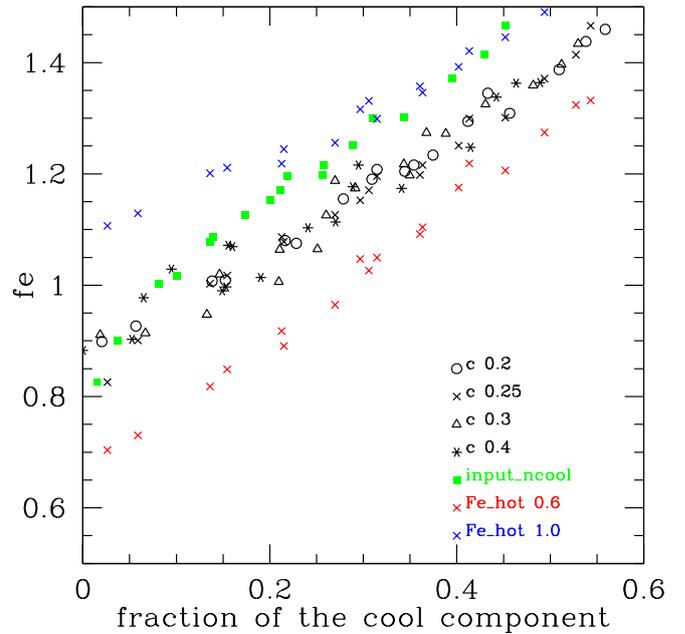}\\
\caption{Results of fitting a simulated combination of two {\it{wdem}} spectra with a high Fe abundance of the cool gas with a {\it{wdem}} model with coupled abundances. Blue, black and red crosses plot the fitted (coupled) Fe abundance against the fitted fraction of the cool component for various input {\it{wdem}} slopes, various upper temperature cutoffs, an input lower temperature cutoff of 0.25$\:T_{\rm{max}}$ and a hot component metallicity of 1.0, 0.8 and 0.6 solar, respectively. Black circles, black triangles and black asterisks plot the same for a hot component metallicity of 0.8 solar and lower temperature cutoffs of 0.2, 0.3 and 0.4, respectively. Green squares plot the fitted (coupled) Fe abundance against the original (input) cool gas fraction for an input lower temperature cutoff of 0.25$\:T_{\rm{max}}$ and a hot component metallicity of 0.8 solar. For an ideal fitting of the cool gas fraction, the black crosses should overlap with the green squares.}
\label{simwdem}
\end{center}
\vspace{-0.5cm}
\end{figure}

The results are presented in Fig. \ref{simwdem}. Most important to note is that from the simulation we do recover, up until quite high percentage contributions of the cool component, a clearly linear dependence of the determined coupled abundance $Z_{\rm{Fe}}^\prime$ on the cool gas fraction $f_{\rm{cool}}^\prime$, plotted on the X axis. The largest scatter is due to the different abundances of the hot gas component, while there is practically no scatter due to $c$ and $T_{\mathrm{max}}$. However, the slope of $Z_{\rm{Fe}}^\prime$ vs. $f_{\rm{cool}}^\prime$ is slightly different than the slope of $Z_{\rm{Fe}}^\prime$ vs. the true $f_{\rm{cool}}$ (by a factor of approximately 1.2).

This discrepancy can be understood in the following way: if the cold gas is more metal-rich than the hot gas and if the spectral model forces the abundances of all gas phases to be the same, then the cool gas in the fitted model will have a too low abundance compared to the real value (and in turn the hot gas will have a too high abundance). To obtain the best fitting spectral shape of the Fe-L complex thus, which is equivalent to obtaining the correct contribution of emission in the FeL energy range from the cold gas, the fit will have to raise the normalization of the cool component to compensate for the fact that the fitted abundance of the cool component is forced to be lower than in reality. This is why coupling the abundances of the different phases leads to an overestimation of the normalization of the cool component, which is reflected in our simulations by a decrease of the $Z_{\rm{Fe}}^\prime$ vs. $f_{\rm{cool}}^\prime$ (fitted) slope compared to the $Z_{\rm{Fe}}^\prime$ vs. $f_{\rm{cool}}$ (real) slope. This affects our estimates of the normalization of the cool gas, as a consequence also our estimates of the mass of the cool gas, and leads to underestimating the metallicity of the cool component if we simply extrapolate the best-fit $Z_{\rm{Fe}}^\prime$ vs $f_{\rm{cool}}^\prime$ line equations to $f_{\rm{cool}}^\prime=1$. If we correct the slopes found in the previous section for the {\it{wdem}} model by the factor described above we find Fe abundances for the cool component as summarized in Table \ref{tab:Fecool}. 

We note that similar simulations as described above were performed also for the two-temperature model, recovering again the expected linear dependence of $Z_{\rm{Fe}}^\prime$ on the cool gas fraction $f_{\rm{cool}}^\prime$, but with a larger scatter due to the assumed metallicities of the hot component. Given the much weaker correlation between $Z_{\rm{Fe}}^\prime$ and $f_{\rm{cool}}^\prime$ in the 2T model and the higher uncertainty in the determination of the observed $Z_{\rm{Fe}}^\prime$ vs. $f_{\rm{cool}}^\prime$ slope, as well as considering our arguments against the two-temperature description of the plasma the X-ray arms, we however choose not to further discuss the 2T results in quantitatively determining $\left< Z_{\rm{Fe_{\rm{cool}}}}\right >$. The assumed thermal model may have an important influence on the determination of the value for $\left< Z_{\rm{Fe_{\rm{cool}}}}\right >$. However, more complicated models cannot be fitted to the available data, since any additional free parameters would be unconstrained. In view of the current lack of a clear theoretical prediction regarding the exact thermal structure of the uplifted gas, the {\it{wdem}} model is at the moment the simplest and most flexible model which serves as a good first-approximation to a wide range of possible emission measure distribution functions. Hence, we will in the following rely on the {\it{wdem}} results for interpretation.

\begin{table}[htdp]
\caption{Estimates of the average iron abundance of the cool component relative to the proto-solar values of \citet{Lodders}.}
\begin{center}
\begin{tabular}{|c|c|c|}
\hline
 & E arm&SW arm\\
  \hline
2T  & 2.38 $\pm$ 0.29 & 2.13 $\pm$ 0.54\\
\hline
wdem uncorrected& 1.93 $\pm$ 0.12 & 1.88 $\pm$ 0.18\\
 \hline
 wdem corrected& 2.16 $\pm$ 0.14 & 2.11 $\pm$ 0.22\\
 \hline
\end{tabular}
\end{center}
\label{tab:Fecool}
\end{table}

Considering the results shown in Table \ref{tab:Fecool}, we will for the rest of this work use an average iron abundance of 2.2 solar. This implies, if we use the abundance ratios presented in Table \ref{abundslopes}, Si and S abundances also around 2.2 solar, and an O abundance of roughly 0.9 solar for the cool gas. Interestingly, the determined Fe abundance for the cool gas is in good agreement with the value of 2.3 solar obtained by \citet{Matsushita07} in the Centaurus cluster, where presumably an undisturbed accumulation of metals over a longer time-scale enabled the central metallicity to reach higher values than in the Virgo cluster center. The difference in the central abundances of M87 and Centaurus can be thus explained by AGN-induced transport of heavy elements out of the central region of M87.

\subsection{The mass of the cool gas}\label{sect:mass}

The emission measure is defined as 
\begin{equation}
Y = \int n_{\mathrm{e}} n_{\mathrm{i}} dV \approx n_{\mathrm{e}} n_{\mathrm{i}} V \approx f_{\rm{ei}}  n_{\mathrm{e}}^2 V
\label{eq:defy}
\end{equation} 
where $n_{\mathrm{e}}$ and $n_{\mathrm{i}}$ are the electron and ion densities, $f_{\rm{ei}}$ is the electron to ion number ratio (on the order of 1.2), and $V$ is the volume of the emitting region. From this, one can estimate the mass of gas according to (where $ \left< m_i \right>$ is the average ion mass per electron in the cluster gas and $m_p$ is the proton mass):
\begin{equation}
M \approx \left< m_i \right> \sqrt{Y V / f_{\rm{ei}}} \approx m_p \sqrt{Y V}
\label{eq:mass}
\end{equation}

This requires assumptions about the projected geometry. To calculate the mass of cool gas in the arms, we compare two alternatives for determining the depth of each bin along the line of sight (LOS). 

\subsubsection{Assumptions about the projected geometry of the cool component}
The first and simplest method is to assume a constant LOS depth for the cool gas component equal to the average width of the radio arms in the plane of the sky. For this, we choose a LOS depth of 200$^{\prime\prime}$ (15.5 kpc). The second possibility is to define the center of the gas halo, determine the minimum and maximum radii of each bin with respect to this center and assume that the longest LOS distance is equal to the longest circle chord that can be drawn between the circle of minimum radius and the circle of maximum radius, in other words that only the gas between the sphere with the minimum radius and the sphere with the maximum radius contribute to the emission \citep[]{Henry04,Mahdavi05}. This yields, where S is the area of the region in the plane of the sky,
\begin{equation}
L=2\sqrt{(R_{\mathrm{max}}^2-R_{\mathrm{min}}^2)}\\{\rm{and}}\\
V=2SL/3.
\label{eq:alexislos}
\end{equation}

We used both of these geometries combined with the integrated emission measure $Y_{\mathrm{cool}}$ of the gas below 1.5 keV from the {\it{wdem}} fit to calculate different estimates of the mass of the uplifted cool gas. In the calculation we included only the bins where the fraction of gas below 1.5 keV was more than $3\sigma$ significant. Both methods show similar spatial distributions of the mass of cool gas. The quantitative results for the total mass of the cool gas are summarized in Table \ref{tab:totmass}, under the column heading ``direct''. For comparison, also the masses of the cool gas determined from the two-temperature model with the two assumed geometries are shown and are in good agreement with the {\it{wdem}} results.

\subsubsection{Filling factor}\label{sect:ff}
For the two volume estimates considered above we additionally must take into account the fact that the cool component may not fill the entire volume of the bin, but may be concentrated in thin filaments \citep{Forman06} which occupy only a fraction (so-called filling factor) of the assumed volume corresponding to each bin. It is very difficult to determine this filling factor from the data, especially since the PSF of XMM-Newton is too large to resolve these filaments. 

The easiest method to estimate the filling factor is to assume pressure equilibrium between the different temperature phases present in the spectral models, which for the {\it{wdem}} model implies $\delta V \propto T^{2+1/\alpha}$ (Sect. \ref{sect:spitzer}), yielding

\begin{equation}
\frac{V_{\rm{c}}}{V_{\rm{h}}}=\frac{\int_{cT_{\rm{max}}}^{1.5} T^{2+1/\alpha}dT}{\int_{1.5}^{T_{\rm{max}}} T^{2+1/\alpha}dT}
\end{equation}

The results for the new mass estimates using the filling factor corrections according to these calculations are presented in Table \ref{tab:totmass} under the column heading ``ff''. We used only the volume estimate given in Eqn \ref{eq:alexislos}, since we cannot assume that the hot component also has a constant line of sight distance of only $200^{\prime\prime}$. These values are however only lower limits to the true mass of the cool gas because the filaments are most likely supported by extra pressure from the magnetic field. If the magnetic field is higher inside the cool component than in the surrounding hot component, then the magnetic pressure adds to the thermal pressure of the cool gas. This reduces the pressure required to maintain hydrostatic equilibrium with the hot surrounding gas, which allows a lower density of the cool gas and a larger filling factor. Note that we have ignored the effects of magnetic tension which can contain the cool gas and work the other way.

To get a more physical estimate of the filling factors we therefore also reevaluated them including the contribution to the pressure of the cold phase of a uniform magnetic field with a strength of 10 $\mu$G, which is a typical average value in the inner radio lobes obtained by \cite{OEK00} using the radio data. The results are likewise shown in Table \ref{tab:totmass}, under the column heading ``ff B10''. We note that the lower limits on the magnetic field strength of 1 $\mu$G obtained in this work (Sect. \ref{sect:pow}) would result in a magnetic pressure several orders of magnitude below the thermal pressure, therefore we did not include this in our calculations but preferred the higher values obtained by \cite{OEK00}.

\subsubsection{Corrections due to coupling the abundances of different temperature phases in the spectral models}\label{sect:corrcoupling}

\begin{figure}[tbp]
\begin{center}
\includegraphics[width=\columnwidth, bb=35 105 577 688]{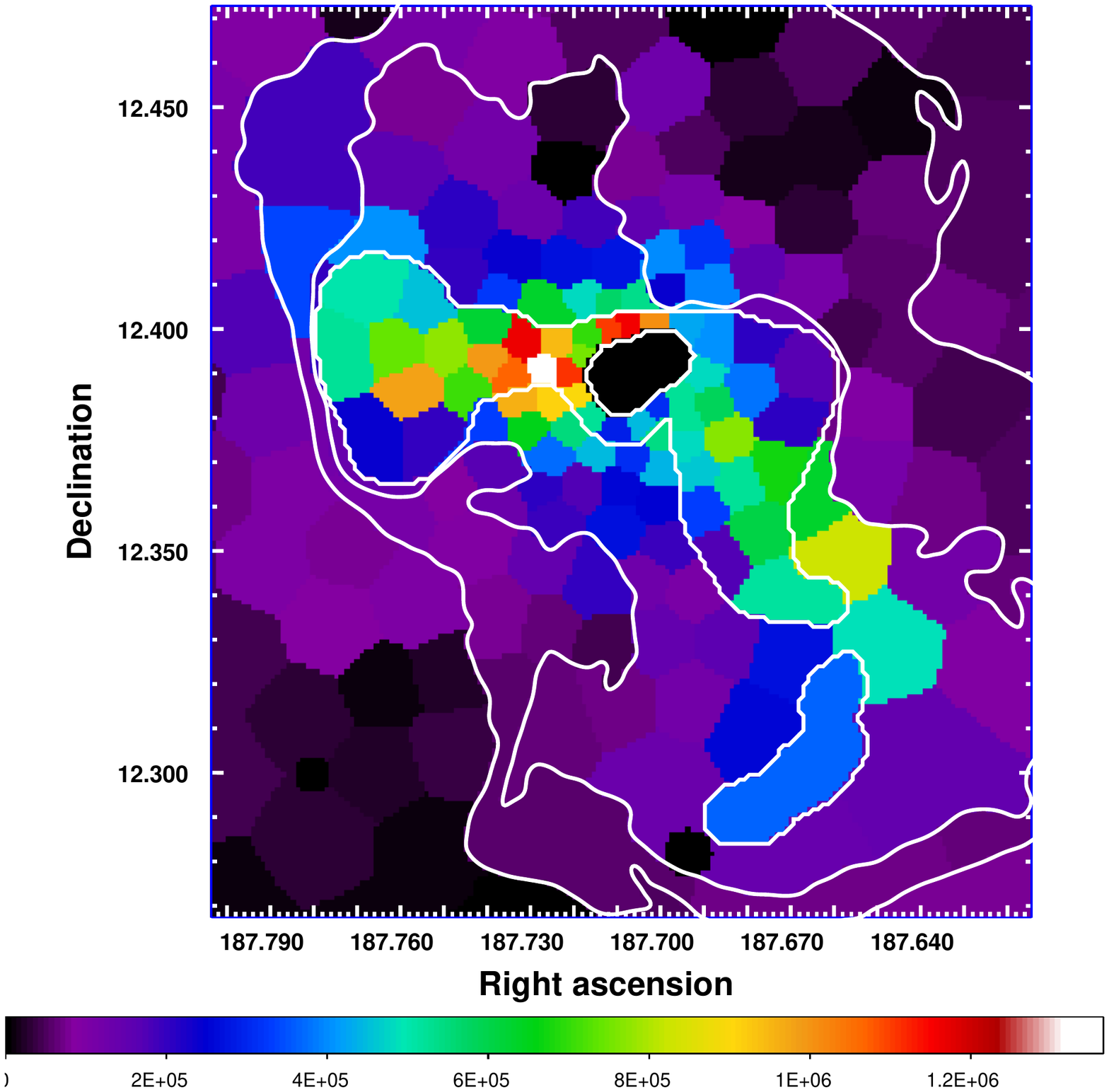}\\
\caption{Map of the mass of gas below 1.5 keV in the {\it{wdem}} model assuming a geometry with a line-of-sight distance for each bin given by Eqn \ref{eq:alexislos}. The map has been corrected for the filling factor of the cool gas assuming pressure support for the cool gas from a 10$\mu$G magnetic field (Sect. \ref{sect:ff}) and for the effects of coupling the abundances of the different phases (Sect. \ref{sect:corrcoupling}). The units are solar masses per square kiloparsec. Radio contours at 90 cm are overplotted in white.}
\label{massmap}
\end{center}
\end{figure}

We proved in the Sect. \ref{sect:sim} that the normalization of the cool gas is overestimated by using a model with a single abundance to describe a combination of a metal-rich component and relatively poorer hot ambient plasma. In the case of a {\it{wdem}} model, this overestimation amounts to a factor of 1.2 if the average metallicity of the cool component is 2 solar (we redid the simulations for an average metallicity of the cool component of 2.2 solar and find that this factor changes by only 5\%).  We therefore further correct the mass of the cool gas decreasing it by this factor, as shown in Table \ref{tab:Fecool} under the column heading ``ffB10 Zcorr''. A map of the mass of cool gas including all the corrections described above is shown in Fig. \ref{massmap}.

\begin{table}[htdp]
\caption{Total mass of the cool gas (in $10^9$ solar masses), including all the different corrections discussed in Sect. \ref{sect:mass}.}
\begin{center}
\begin{tabular}{|c|c|c|c|c|c|c|c|}
\hline
 LOS&model&direct&ff&ff&ffB10\\
 & & & &B10&Zcorr\\
 \hline
 const&2T&2.3&-&-&-\\
 \hline
 Eqn \ref{eq:alexislos}&2T&2.4&0.39&0.41&-\\
 \hline
 const&{\it{wdem}}&2.2&-&-&-\\
 \hline
 Eqn \ref{eq:alexislos}&{\it{wdem}}&2.1&0.54&0.60&0.50\\
 \hline
\end{tabular}
\end{center}
\label{tab:totmass}
\vspace{-0.5cm}
\end{table}

In the following, we will use the estimate of $5 \times 10^8 M_\odot$ for the mass of the cool gas, which includes all the corrections discussed in this section and is determined from the {\it{wdem}} model which we believe to describe the data better than the two-temperature model.
We note that  the cooling time of the cool gas in the multi-temperature regions, computed assuming isobaric cooling and using an average value of the temperature for the multiphase distribution, is longer than the ages of the radio lobes in all bins (between $2 \times 10^8$ and $2 \times 10^9$ years). Therefore our mass estimate is not affected by gas which has cooled out of the X-ray domain since the uplift.

\subsection{Discussion of the origin and uplift of the cool gas}

We have presented so far in this work many arguments in favor of the scenario proposed by \cite{Churazov01}, in which the cold gas in the X-ray arms is a result of gas uplift from the central regions of the galaxy through buoyant radio bubbles. In this section we will try to bring the observational details together towards a consistent picture of the energetics and chemical composition of the M87 X-ray arms and explain how this relates to the radio activity.

\subsubsection{Energetics}

We first use the mass estimates derived in the previous section to compute the gravitational energy needed to uplift the gas:
\begin{equation}
U = \frac{GM_{\mathrm{cool}}M(R)}{R}
\end{equation}
For M(R) we use the integrated total mass profile of \cite{Matsushita02}. We use the approximation that all the gas with total mass $M_{\rm{cool}}$ is uplifted from R=0 and U=0 to a mean radius 
\begin{equation}
R_{\mathrm{mean}}=\frac{ \sum M(i) R(i)}{M_{\mathrm{cool}}}
\end{equation}
where i are all the bins where the cool component was more than 3$\sigma$ significant. We calculated $R_{\mathrm{mean}}$ in this way for all the different models and volume estimates presented in the previous section and obtain consistent values between 15 and 20 kpc. 
Assuming a typical mass of the uplifted gas of $0.5\times10^9\:M_\odot$ (see Table \ref{tab:totmass}) and $R_{\mathrm{mean}}$ = 20 kpc, we obtain a required gravitational energy of $4.33\times10^{57}$ ergs. If the AGN outburst that created the inner radio lobes occurred $10^7$ years ago \citep[e.g.][]{Forman06}, this gravitational energy corresponds to a required power of $1.37\times10^{43}$ erg/s, which is a sizable fraction of the total jet power estimated by \cite{OEK00} at a few $\times 10^{44}$ ergs/s. The power needed to uplift the cool gas is much larger than the total radio power cited by \cite{OEK00} at only $9.6\times10^{41}$ erg/s, consistent with the picture that the bulk of the jet energy is deposited in the gas (in this case, a large part thereof in the form of potential energy), while only a small fraction is leaked to radio emission. 
 
Thus, from energetic considerations, the model of the cool gas being uplifted by the buoyant radio lobes is entirely consistent with the X-ray and radio observations. This provides a clear mechanism for the transport of the cool gas out of the center of the gas halo, together with the metals it has been enriched with. 

\subsubsection{Possible sources of cool gas}

We have chosen to define the cool gas in the {\it{wdem}} model as gas having a temperature below 1.5 keV, the smallest value in the deprojected profile of the hot M87 halo calculated by \citet{Matsushita02}. There are three possible ways to produce gas cooler than 1.5 keV. The most straight-forward option is to adiabatically cool gas from the center of the M87 halo by uplifting it into regions of lower pressure at larger radii. Using the pressure profile of \citet{Simionescu07} and assuming an adiabatic index of 5/3, gas at 1.5 keV uplifted from a radius of 1 to 10 kpc would cool from 1.5 to 0.6 keV and so could reproduce, in principle, the temperatures that we observe. However, if adiabatic cooling is the only mechanism to reduce the temperature of the gas, we would (neglecting heating processes) expect to see a decreasing trend of $cT_{\rm max}$ with radius, which is not the case. More importantly, the average Fe abundance of 2.2 solar for the cool gas, determined in Sect. \ref{sect:sim}, is higher than the value observed in the center of M87 \citep[$\sim$1.6 solar,][]{Matsushita03}. Thus, the cool gas could not originate from adiabatically cooled halo gas alone, but must contain a second, metal-rich component.

Another important source of cool gas are the stellar winds. Assuming the stellar mass loss rate proposed by \cite{Ciotti91},
\begin{equation}\label{massloss}
\dot{M}_*(t) \approx -1.5\times10^{-11}L_B\:t_{15}^{-1.3}\:,
\end{equation}
where $L_B$ is the blue band luminosity in units of $L_{B\odot}$ and $t_{15}$ is the age of the stellar population in units of 15 Gyrs. We take the age of the stellar population to be 10 Gyrs and $L_B$ to be $10^{11}L_{B\odot}$, corresponding to the absolute B magnitude of M87 of -22.14 cited by \cite{Peletier90}. This gives us a stellar mass loss rate of 2.54 $M_\odot$/yr. If we further assume that the material lost from stars thermalizes according to the M87 stellar velocity dispersion of 350 km/s \citep{Angione80}, the temperature of this gas would be 0.64 keV. Interestingly, this is also very close to the lower temperature cutoffs found using the {\it{wdem}} spectral model. Beside being a source of cool gas, stellar winds also contribute to the metal-enrichment, as will be discussed in the next section, and are an important candidate for the second, metal-rich component of the uplifted gas.

The third and last possible source of cool gas is the cooling flow. \cite{Matsushita02} find that within $4^\prime$ a total mass of 13.8 $M_\odot$/yr can cool down to as low as 1.4 keV and 0.63 $M_\odot$/yr cool down to 0.1 keV. Based on the same metallicity argument as above, the cool gas could not originate from cooling-flow gas alone since this would lead to its metallicity being lower than measured. Using the gas metallicity and temperature, it is not possible to distinguish adiabatically and radiatively cooled ICM as different components of the uplifted gas.

\subsubsection{Metal enrichment of the cool gas}

Metals can be either ejected from the stars by the stellar winds or produced by supernovae. We neglect the metal enrichment by type II supernovae, since M87 has an old population where the expected SNII rate is very low. \citet{Rafferty06}, for example, find an upper limit on the star formation rate (SFR) for M87 of 0.081 $M_\odot/\rm{yr}$ using far-infrared data. Assuming a Salpeter initial mass function \citep[IMF,][]{Salpeter55} for the stars forming between 0.01 and 50 solar masses, the number of stars formed from this cooling gas that would be more massive than 10 solar masses and could become SNII progenitors is
\begin{equation}
N(>10M_\odot)=0.081\cdot\frac{\int_{10}^{50}M^{-2.35}dM}{\int_{0.01}^{50}M\:M^{-2.35}dM}\approx 0.017/\rm{century}
\end{equation}
This upper limit on the rate of SNeII is roughly two orders of magnitude below the expected rate of SNeIa in elliptical galaxies \citep[][ see calculation below]{Cappellaro99}.

This leaves stellar winds and type Ia supernovae (SN~Ia) as the main sources of enrichment of the cool gas. Following \citet{Boehringer04,Rebusco06}, the rates of iron injection by stellar mass loss is:
\begin{eqnarray}\label{eq:feratestar}
\dot M_{\rm{Fe,*}} & = & \gamma_{\rm Fe}\times \dot{M}_*(t), {\rm with}\\
\gamma_{\rm Fe} & = & Z_{\rm Fe,*}\times \frac{\left< m_{\rm Fe}\right> }{\left<m_{i\odot}\right>} \times \left(\frac{N_{\rm Fe}}{\sum_i N_i}\right)_{\rm Lodd}, \label{gamma_Fe}
\end{eqnarray}
where $\dot{M}_*(t)$ is the stellar mass loss rate calculated in Eqn. \ref{massloss}, $\gamma_{\rm Fe}$ is the mean iron mass fraction in the stellar winds of an evolved stellar population, $Z_{\rm Fe,*}$ is the assumed abundance of the stellar winds in solar units, $\left<m_{\rm Fe}\right>\approx56$ is the mean weight of an iron isotope with respect to hydrogen, $\left<m_{i\odot}\right>$ is the mean weight of an ion with respect to hydrogen in the Sun (in the solar photosphere), and $ \left(N_{\rm Fe}/\sum_i N_i\right)_{\rm Lodd}\approx3.16\times10^{-5}$ is the relative number abundance of Fe ions to the sum over all ions in the plasma, as determined from \citet{Lodders}. We assume, as above, a galactic age of 10 Gyr. \citet{Kobayashi99} derive stellar metallicities in M87 ranging from $\approx$2 solar in the center to $\approx$1 solar at the galaxy's half-light radius ($2^\prime$) using the $\rm Mg_2$ index. We will therefore take $Z_{\rm Fe,*}\approx1.6$. The rate of iron production by SN~Ia can be written as:
\begin{eqnarray}
\dot M_{\rm{Fe,Ia}}& = & \eta_{\rm Fe} \times R_{Ia} \rm{, with} \label{eq:ferateIa}\\
R_{Ia} & = &  0.18\pm0.06  \: ({\rm{100yr}})^{-1} \left(10^{10}L_{B\odot} \right)^{-1}
\label{eq:rateIa}
\end{eqnarray} 
where $\eta_{Fe}=0.79 M_{\odot}$ is the iron yield per SNIa assuming the WDD2 model of \citep{Iwamoto99} and $R_{Ia}$ is the present supernova rate in elliptical galaxies according to \citet{Cappellaro99}. For M87, $L_B\approx10^{11}L_{B\odot}$, as calculated above, which yields $R_{Ia} \approx 1.8 \pm 0.6$ SNeIa per century. We assume that the iron production by both SN~Ia and stellar winds is constant during the time of enrichment, which as we will show below is much shorter that the Hubble time.

We wish to enrich ICM gas to an average Fe abundance of 2.2 solar. For this, we must assume an initial metallicity $Z_{\rm i}$ of the ICM at the center of M87 immediately following the next to last AGN outburst which modified the spatial distribution of metal abundances. After the next to last AGN outburst, stellar mass loss and supernovae enriched this central gas until it reached 2.2 solar and was uplifted by the last AGN outburst, the effect of which we observe today. Ideally we should take the central metallicity immediately after the last outburst and assume it is representative also for the metallicity after the next to last outburst. We however do not know this value. A sensible lower estimate is to assume $Z_{\rm i}$ as the average metallicity of the entire hot halo within the region of the arms. This can be calculated as the best-fit y-intercept for the linear fit of the Fe abundance in the multi-temperature regions vs. the fraction of cool gas (Fig. \ref{percool_vs_fe}) and is approximately 0.7 solar. Alternatively, as an upper limit to $Z_{\rm i}$, we can assume the current value at the center of M87 determined by \citet{Matsushita03} to be 1.6 solar. This is likely to be an upper limit because, since the time of the last outburst, the central region has been enriched with metals beyond the metallicity that it had immediately after the outburst. With these two values for $Z_{\rm i}$, we can bracket the enrichment time needed to reproduce the observed Fe abundance of the cool gas.

Using a mass production rate of stellar winds of 2.5 $M_\odot/\rm{yr}$ (Eqn. \ref{massloss}) and neglecting the mass of supernova remnants (assuming an average supernova mass of 10 $M_{\odot}$ and rate $R_{SN~Ia}$ of 1.8 per century, we obtain only 0.18 $M_\odot/\rm{yr}$ which is much less than for the stellar winds), we can write for the Fe enrichment of the uplifted material:

\begin{eqnarray}
2.2\times5\cdot10^8M_{\odot}\times f_{\rm{Fe},\odot}&=&Z_{\rm i}\times M_{\rm i} \times f_{\rm{Fe},\odot} \\
&+&(\dot M_{\rm{Fe,Ia}}+\dot M_{\rm{Fe,*}})\times\tau\nonumber
\end{eqnarray}

\begin{equation}
M_{\rm i}+2.5 M_\odot/\rm{yr} \times \tau = 5\cdot 10^8 M_\odot
\end{equation}
where $f_{\rm{Fe}\odot}= \left< m_{\rm Fe}/m_{i\odot} \right> \times \left(N_{\rm Fe}/\sum_i N_i\right)_{\rm Lodd}$ is the mass fraction of Fe in the Sun, $\tau$ is the time over which the enrichment takes place, $M_{\rm i}$ is the initial mass of cluster gas to which stellar mass loss is added over time $\tau$. We solve the system of equations above for $\tau$ and $M_{\rm i}$ making use also of Eqns. \ref{eq:feratestar} and \ref{eq:ferateIa}. Assuming $Z_{\rm i}$=1.6 solar we obtain $\tau$=30 Myr, while for $Z_{\rm i}$=0.7 solar we need 60 Myr to enrich the ICM before uplift. For $Z_{\rm i}$=1.6 solar, the mass originating from stellar mass loss represents 15\% of the total uplifted mass, while for $Z_{\rm i}$=0.7 solar it amounts to 30\%.

We can now try to reproduce the abundances of other elements such as O, Si and S in the cool gas. Using the Fe/O, Fe/Si, Fe/S ratios in Table \ref{abundslopes}, the O, Si and S abundances of the cool gas should be approximately 0.9, 2.2 and 2.2, respectively. For an initial Fe abundance of $Z_{\rm i} = 0.7$ solar and assuming the same ratios, we obtain 0.3, 0.7 and 0.7 solar for O, Si and S, respectively. The central values of \citet{Matsushita03} are consistent with 0.6,1.6, and 1.6, respectively. 

Assuming solar abundance ratios for the stellar winds, we can use the equivalent of Eqn. \ref{eq:feratestar} and \ref{gamma_Fe} to determine also $\dot M_{\rm{O,*}}$. Using $Z_{\rm O,i}$=0.6 and ignoring contributions by SN~Ia to the oxygen budget, the time needed to enrich the cluster gas to 0.9 solar oxygen abundance is 60 Myr. Correspondingly for $Z_{\rm O,i}$=0.3, we obtain 92 Myr. A similar procedure can be applied for Si and S, including contributions by SN~Ia according to the WDD2 model of 0.21 $M_\odot$ and 0.12 $M_\odot$ per supernova, respectively. For Si we obtain, respectively, 63 and 108 Myr for $Z_{\rm Si,i}$=1.6 and 0.7, while for S we obtain 56 and 98 Myr for the same $Z_{\rm S,i}$.

We also computed the required enrichment times using the WDD1 model of \citet{Iwamoto99} which has a higher Si/Fe ratio and was found by \citet{Finoguenov02} to provide a better fit to the abundance patterns in the center of the M87 gas halo. These results, together with the results from the WDD2 model, are summarized in Table \ref{tab:enrichtime}. Oxygen enrichment times for the two models are the same since we neglect contributions by SN~Ia to the oxygen budget.
The WDD1 model requires more similar enrichment times to produce the different metals, in agreement with the results of \citet{Finoguenov02}. The obtained enrichment times differ from eachother by up to a factor of $\approx 2.1$ for the WDD2 and $\approx 1.7$ for the WDD1 models. A possible source of this discrepancy is that our assumption of solar abundance ratios for the stellar winds is not accurate. If, for the older population of M87, the Si/Fe composition of stellar winds would be higher than solar, we would obtain more consistent enrichment times to produce all three metals. Note also that the enrichment times are in slightly better agreement for $Z_{\rm i} = 0.7$ solar than for $Z_{\rm i} = 1.6$ solar, suggesting that the former choice may be the more appropriate and reliable one.

\begin{table}[htdp]
\caption{Enrichment times needed to produce the metals seen in the cool gas (in Myr) and, in brackets, the fraction of the stellar mass loss contribution to the total mass of uplifted gas.}
\begin{center}
\begin{tabular}{|l|l|l|l|l|}
\hline
 &\multicolumn{2}{l|}{$Z_{\rm Fe/Si/S,i}=1.6$}&\multicolumn{2}{l|}{$Z_{\rm Fe/Si/S,i}=0.7$}\\
 &\multicolumn{2}{l|}{$Z_{\rm O,i}=0.6$}&\multicolumn{2}{l|}{$Z_{\rm O,i}=0.3$}\\
\cline{2-5}
 &WDD2&WDD1&WDD2&WDD1\\
\hline
Fe & 30 (0.15) & 35 (0.18)& 60 (0.30)& 69 (0.35)\\
Si & 63 (0.32) & 49 (0.25)& 108 (0.54)& 90 (0.45)\\
S  & 56 (0.28) & 42 (0.21)&98 (0.49)& 79 (0.40)\\
O & 60 (0.30) & 60 (0.30)& 92 (0.46)& 92 (0.46) \\
\hline
\end{tabular}
\end{center}
\label{tab:enrichtime}
\end{table}

At the end of Sect. \ref{sect:metalpatterns}, we show that the relative abundances of O/Si/ S/Fe are very similar in and outside the multi-temperature regions. This indicates firstly that the metals outside the multi- temperature region, which were (according to our model) uplifted from the galaxy by previous AGN outbursts, must have been transported after the last major epoch of star formation, when contributions of SNII to the chemical budget of the galactic gas became negligible. Secondly, the SN~Ia enrichment must have remained approximately constant in time relative to the enrichment by stellar winds.

We note that \cite{OEK00} estimate the age of the outer radio halo between 96 and 150 Myr/$P_{44}$, where $P_{44}$ is the total jet power in $10^{44}$ erg/s. If the total jet power is around 1--3 $\times 10^{44}$ erg/s, we would obtain consistent values with the enrichment times of $\sim$30--110 Myr obtained above. Thus, it is possible that both the outer and the inner radio halos generated similar outbursts $\sim$ 3--11$\times10^7$ and $\sim10^7$ years ago respectively, transporting metal-rich gas outwards, while between the two outbursts the metals could accumulate in the center of the gas halo. Less violent outbursts which do not have such large impact in the distribution of metals in the gas halo could have occurred between these two events, as is suggested by the multiple shocks found in the Chandra data by \cite{Forman06}. 

Note that if some of the uplifted metal-rich gas fell back to where it originated between outbursts, the time interval between two such outbursts could be shorter since one would not have to wait for all the extra metals observed in the cool gas to be produced from stellar winds and supernovae.
%________________________________________________________________

\section{Constraints on possible non-thermal emission}\label{sect:pow}
%________________________________________________________________

\begin{figure}[tbp]
\begin{center}
\includegraphics[width=\columnwidth, bb=36 104 576 689]{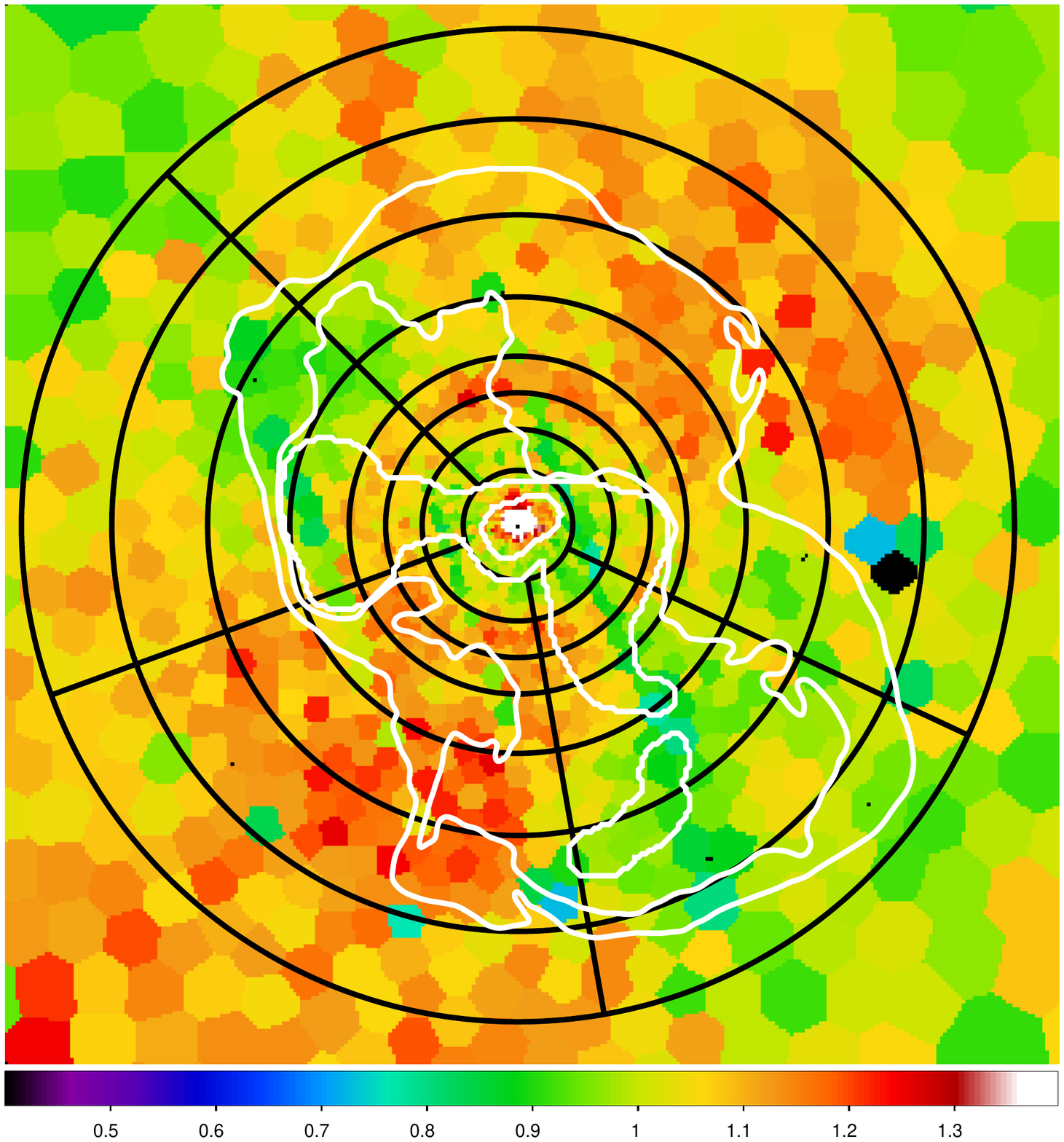}\\
\caption{Regions for spectral extraction used to test the presence of a power-law component (in black) overplotted on the map of the gas pressure deviations from radial symmetry \citep{Simionescu07}. 90cm radio contours are also overplotted in white.}
\label{extractionrings}
\end{center}
\end{figure}

\cite{Simionescu07} suggest the presence of non-thermal pressure support by relativistic electrons in the X-ray arms of M87.  We performed a more detailed analysis to check for spectral signatures of inverse Compton (IC) scattering of the microwave background photons on the relativistic electrons in the M87 gas halo. For this, we chose 7 annuli between 1 and 9 arcminutes and divided each of these into 4 sectors containing the E arm, the SW arm, and the NW and SE regions between the arms. The sectors containing the arms were fitted with a two-temperature VMEKAL model and a power-law while the NW and SE regions were fitted with a single-temperature VMEKAL model plus a power-law. Our extraction regions can be seen in Fig. \ref{extractionrings}.

We find that the fit cannot constrain the power-law index in any of the extraction regions, therefore we fixed it to 1.7, which is consistent with the number distribution of relativistic electrons in our Galaxy. With a fixed power-law index, we find that the normalization of the power-law component is more than $3\sigma$ significant everywhere except in the 7th ring of the SW arm sector. The flux of the power-law component in all regions is a factor of more than 10 above the cosmic X-ray background (CXB) found by \cite{DeLuca04} in the 2-10 keV band and a factor of more than 5 above the same CXB model in the 7-10 keV band. However, we cannot rule out the possibility that the power-law component is significant simply because it compensates for calibration or spectral model inaccuracies, therefore we cannot confirm it as a detection of IC emission from the M87 halo, but interpret our results rather as an upper limit of the IC flux that is consistent with the data. 

For the upper limit on the non-thermal pressure, we use \citep{Lieu99}
\begin{eqnarray}
p=\frac{E}{3V} \:\:\:\rm{where}\\
E=8\times10^{61}\rm L_{42}\times \frac{3-\mu}{2-\mu}\times\frac{\gamma_{max}^{2-\mu}-\gamma_{min}^{2-\mu}}{\gamma_{max}^{3-\mu}-\gamma_{min}^{3-\mu}}\:\:\:\rm{and}\\
V=\frac{2\pi}{3}(1-\rm{cos}\:\theta)(R_{\rm{max}}^3-R_{\rm{min}}^3).\label{eqn:vsphshell}
\end{eqnarray}
$L_{42}$ is the luminosity of the non-thermal emission between $\gamma_{\rm{max}}$ and $\gamma_{\rm{min}}$ in units of $10^{42}$ ergs, $\mu$ is the index of the electron number distribution which is related to the photon index of our power-law model as $\mu=2\Gamma-1=2.4$, $\gamma_{\rm{min}}$ and $\gamma_{\rm{max}}$ are the minimum and maximum Lorentz factors of the electrons contributing to the IC emission, V is the volume of a spherical shell with half-opening angle $\theta$ and minimum and maximum radii $R_{\rm{min}}$ and $R_{\rm{max}}$. We extrapolated our power-law luminosity to include IC scattering by electrons with Lorentz factors between $\gamma_{\rm{min}}=300$ and $\gamma_{\rm{max}}=10000$ and find 90\% confidence upper limits on the non-thermal pressure as presented in Table \ref{tab:pnth}. Note that the pressure calculated in this way is only due to the relativistic electrons, and that relativistic ions may give rise to additional non-thermal pressure up to 30 times higher than that of the electrons. We also computed the thermal pressure as $p=(1+1/f_{ei})n_ekT$ with $n_e$ given by Eqn \ref{eq:defy} in order to present also the relative contribution of the non-thermal pressure to the total (thermal+non-thermal) pressure in Table \ref{tab:pnth}.

\begin{table}[htdp]
\caption{90\% confidence level upper limits on the non-thermal pressure (in $10^{-11}\: \rm{dyn/cm^2}$ and in brackets as percentage of the total pressure).}
\begin{center}
\begin{tabular}{|c|c|c|c|c|}
\hline
 $ r_{\mathrm{mean}}$&NW &SE&E &SW \\
(arcmin)& & &arm&arm\\
 \hline
1.38&7.4 (31.5) &27.6 (54.4)&18.2 (40.9)&19.2 (40.5)\\
 \hline
2.09&3.2 (19.8)&7.9 (30.1)&6.7 (25.7)&8.5 (32.2)\\
 \hline
2.75&1.7 (14.1)&2.8 (16.5)&4.5 (22.6)&2.3 (13.4)\\
 \hline
3.63&0.85 (9.4)&1.2 (10.3)&2.1(17.3)&2.1(15.9)\\
 \hline
4.92&0.68 (9.9)&0.50 (5.5)&1.1 (12.6)&1.3 (13.6)\\
 \hline
6.55&0.22 (4.4)&0.42 (6.1)&0.53 (7.5)&0.84 (11.3)\\
 \hline
8.25&0.14 (3.5) &0.22 (4.1)&0.31 (5.7)&0.26 (4.5)\\
 \hline
\end{tabular}
\end{center}
\label{tab:pnth}
\end{table}

The upper limits presented in Table \ref{tab:pnth} are consistent with the contribution of the non-thermal pressure of 5-10\% suggested by \cite{Simionescu07}. However, from these upper limits no difference between the arm and off-arm regions can be seen, suggesting that perhaps at least partly the normalization of the power-law is determined by calibration uncertainties.
The high contribution of the power-law in the inner regions is probably due to the central AGN, which was very bright during this observation and could influence the spectrum out to larger radii due to the tail of the XMM PSF.

We also used our upper limits on the power-law flux, assuming it is due to IC emission, to determine the lower limits on the magnetic field stength, which can be obtained from the equation \citep{Sarazin86}
\begin{eqnarray}
\frac{f_{x}}{f_{r}}\left(\frac{\nu_x}{\nu_r}\right)^{\alpha_x}=\frac{2.47\times10^{-19}T_r^3b(\mu)}{B\:a(\mu)}\times\left(\frac{4960T_r}{B}\right)^{\alpha_x}
\end{eqnarray}
where $a(\mu)$ and $b(\mu)$ are unitless functions of $\mu$ only, and for our case $\mu=2.4$, $a=0.086$ and $b=7.0$, $\alpha_x=(\mu-1)/2$ is the X-ray spectral index, $T_r$ is the temperature of the microwave background radiation (2.73 K), B is the magnetic field strength, $f_x$ is the X-ray flux density at frequency $\nu_x$ in ergs/s/$\mathrm{cm}^2$/Hz and $f_r$ is the radio flux density at frequency $\nu_r$. We used the radio map of \cite{OEK00} at 90 cm ($\nu_r$=327 MHz) and chose to calculate the X-ray flux at 6 keV by dividing the 5 - 7 keV power-law flux by the corresponding bandwidth in Hz. Our lower limits on the magnetic field strength B are summarized in Table \ref{tab:b}. We could only determine the values for B in the first 6 rings (5 for the E arm) because the other regions were outside the radio map.
\begin{table}[htdp]
\caption{90\% confidence level lower limits on the magnetic field strength ($\mu$G). The sixth ring in the E arm was outside the radio map}
\begin{center}
\begin{tabular}{|c|c|c|c|c|}
\hline
 $ r_{mean}$&NW &SE&E &SW \\
(arcmin)& & &arm&arm\\
 \hline
1.375&0.53 &0.31& 0.51&0.65\\
 \hline
2.085&0.62&0.48&0.77&0.94\\
 \hline
2.750&0.56&0.64&0.96&1.40\\
 \hline
3.625&0.60&0.78&1.26&1.09\\
 \hline
4.920&0.47&0.82&0.52&1.16\\
 \hline
6.545&0.29&0.52&-&1.20\\
 \hline
\end{tabular}
\end{center}
\label{tab:b}
\end{table}

Interestingly, we do find more stringent lower limits on the magnetic field in the X-ray arms, these being the only regions where our data require magnetic fields stronger than 1 $\mu$G. However, as these values are only lower limits, we cannot draw definitive conclusions based on this fact. 

Assuming equipartition, \cite{OEK00} find B-field strengths on the order of 7 - 10 $\mu$G, which are consistent with our lower limits. However, in cluster cores, where the particles and magnetic fields have different origins and evolutionary histories, the validity of the equipartition condition is not obvious. To measure a magnetic field of 10 $\mu$G using X-ray observations would require the ability to determine with significance a power-law flux 50 times lower than our upper limit, which, within the current calibration accuracies of $\sim$10\% is still out of reach. Using the Faraday rotation of the halo radio source in M87 \citet{dennison1980} finds a magnetic field in the halo of 2.5 $\mu$G. 
While the magnetic field deduced using the Faraday rotation measure is the average along the line of sight of the product of the magnetic field and the gas density, the lower limit of the magnetic field determined here depends on the volume average of the relativistic electron density and on the square of the magnetic field over the emitting region. The fields determined using these two methods can be quite different, and as shown by \citet{goldshmidt1993} the magnetic field determined using the radio and IC X-ray flux densities
is in general smaller than the value determined using the Faraday rotation measure.

\citet{sanders2005} found a non-thermal X-ray emission component in the core of the Perseus cluster extending to a radius of $\sim$75 kpc. Assuming this emission to be due to IC scattering of the cosmic microwave background and infrared emission from NGC~1275, they mapped the magnetic field in the core of the cluster. Within the radius of $\sim$10~kpc they found a magnetic field between 0.5--3 $\mu$G, while at larger radii their inferred magnetic field decreased to a value of $\sim$0.1$\mu$G. The higher values in the core are mostly consistent with the lower limits found in the core of M87, but the inferred magnetic field at larger radii is significantly smaller than the lower limits we found at similar radii in M87. 

For the radio relic in Coma, combining the radio data with XMM-Newton observations, \citet{feretti2006} found a lower limit on the magnetic field of 1.05~$\mu$G, which is consistent with our lower limits found for the radio arms. The deduced magnetic field in Coma based on the detection of tails of hard X-ray emission by BeppoSAX \citep{fusco-femiano2004} is $\sim$0.2~$\mu$G. However, the cluster region probed by the large field of view of the BeppoSAX PDS is much larger than the core of Virgo investigated in this paper.

\section{Conclusions}
\label{sect:conclusions}
We used deep XMM-Newton observations of the M87 halo to characterize the spatially resolved temperature structure and the chemical composition of the multi-temperature gas associated with the inner radio lobes. We found that:

\begin{itemize}

\item
Compared to a simple two-temperature fit, we obtain a better and more physical description of the spectra using a model which involves a continuous range of temperatures in each spatial bin. The range of temperatures of the multiphase gas spans between $\sim$0.6--3.2 keV. The cooling time of this gas is longer than the age of the lobes, indicating that no significant amounts of gas cooled out from the temperature domain observable in X-rays.  If the multi-phase gas is distributed in many small spherical blobs with typical blob sizes of 0.1 kpc, similar to the widths of the filaments seen in the Chandra images \citep{Forman06}, the thermal conduction must be suppressed by a factor of $\sim$30--100. Such suppression may be  provided by the magnetic fields.

\item
We find a correlation between the amount of gas cooler than 1.5~keV and the metallicity, indicating that the cool gas is more metal-rich than the ambient halo. Extrapolating the linear fit between the cool gas fraction and Fe abundance, we estimate the Fe abundance of the cool gas to $\sim$2.2 solar. This suggests the key role of the AGN in transporting heavy elements into the intracluster medium by uplifting cool, metal-rich gas from the galaxy.
The abundance ratios of O/Si/S/Fe in and outside the X-ray arms are similar, indicating that the metals outside the multi-temperature region, which were (according to our model) uplifted from the galaxy by previous AGN outbursts, must have been transported after the last major epoch of star formation, when contributions of SNII to the chemical budget of the galactic gas became negligible. 

\item
We estimate the mass of the gas with a temperature below 1.5 keV (the smallest value in the deprojected profile of the hot M87 halo) to be about $5\times10^{8}~M_{\odot}$. Its chemical composition indicates that it originates from a mixture of ICM and stellar mass loss enriched with Type Ia supernova products. The amount of mass from stellar winds and SNeIa represents 15--50\% of the total mass of uplifted gas. The time required to produce the observed metals in this gas is $\approx$30--110 Myr, suggesting that the uplift of cool gas by AGN radio bubbles is a relatively rare event. 

\item
We put upper limits on possible non-thermal X-ray emission from M87, and combining it with the 90~cm radio maps, we put lower limits of around 0.5-1.0~$\mu$G on the magnetic field strength.

\end{itemize}

%________________________________________________________________
\begin{acknowledgements}
We would like to thank E. Churazov and A. Baldi for very helpful discussions. We also thank an anonymous referee for constructive suggestions. We acknowledge the support by the DFG grant BR 2026/3 within the Priority Programme "Witnesses of Cosmic History". 
NW acknowledges the support by the Marie Curie EARA Early Stage Training visiting fellowship. AF acknowledges support by NASA grants NNX06AE79G, NNG05GM50G and NNG05GH40G, as well as grant 50 OR 0207 from BMBF/DLR.
The XMM-Newton project is an ESA Science Mission with instruments and contributions directly funded by ESA Member States and the USA (NASA). The XMM-Newton project is supported by the Bundesministerium f\"ur Wirtschaft und Technologie/ Deutsches Zentrum f\"ur Luft- und Raumfahrt (BMWI/DLR, FKZ 50 OX0001), the Max Planck Society, and also by PPARC, CEA, CNES, and ASI. The Netherlands Institute for Space Research (SRON) is supported financially by NWO, the Netherlands Organization for Scientific Research. 
\end{acknowledgements}

\bibliographystyle{aa}
\bibliography{bibliography}

\begin{thebibliography}{64}
\expandafter\ifx\csname natexlab\endcsname\relax\def\natexlab#1{#1}\fi
\expandafter\ifx\csname url\endcsname\relax
  \def\url#1{{\tt #1}}\fi
\expandafter\ifx\csname urlprefix\endcsname\relax\def\urlprefix{URL }\fi

\bibitem[{{Akritas} \& {Bershady}(1996)}]{Akritas96}
{Akritas} M.G., {Bershady} M.A., Oct. 1996, \apj, 470, 706

\bibitem[{{Anders} \& {Grevesse}(1989)}]{Angr}
{Anders} E., {Grevesse} N., Jan. 1989, \gca, 53, 197

\bibitem[{{Angione} et~al.(1980){Angione}, {Junkkarinen}, {Talbert}, \&
  {Brandt}}]{Angione80}
{Angione} R.J., {Junkkarinen} V., {Talbert} F.D., {Brandt} J.C., Apr. 1980,
  \pasp, 92, 149

\bibitem[{{Arnaud} \& {Raymond}(1992)}]{Arnaud92}
{Arnaud} M., {Raymond} J., Oct. 1992, \apj, 398, 394

\bibitem[{{Belsole} et~al.(2001){Belsole}, {Sauvageot}, {B{\"o}hringer}
  et~al.}]{Belsole01}
{Belsole} E., {Sauvageot} J.L., {B{\"o}hringer} H., et~al., Jan. 2001, \aap,
  365, L188

\bibitem[{{Binney} \& {Tabor}(1995)}]{Binney95}
{Binney} J., {Tabor} G., Sep. 1995, \mnras, 276, 663

\bibitem[{{B{\^i}rzan} et~al.(2004){B{\^i}rzan}, {Rafferty}, {McNamara},
  {Wise}, \& {Nulsen}}]{Birzan04}
{B{\^i}rzan} L., {Rafferty} D.A., {McNamara} B.R., {Wise} M.W., {Nulsen}
  P.E.J., Jun. 2004, \apj, 607, 800

\bibitem[{{B\"ohringer} \& {Fabian}(1989)}]{Boehringer89}
{B\"ohringer} H., {Fabian} A.C., Apr. 1989, \mnras, 237, 1147

\bibitem[{{B\"ohringer} et~al.(1995){B\"ohringer}, {Nulsen}, {Braun}, \&
  {Fabian}}]{Boehringer95}
{B\"ohringer} H., {Nulsen} P.E.J., {Braun} R., {Fabian} A.C., Jun. 1995,
  \mnras, 274, L67

\bibitem[{{B{\"o}hringer} et~al.(2002){B{\"o}hringer}, {Matsushita},
  {Churazov}, {Ikebe}, \& {Chen}}]{Boehringer02}
{B{\"o}hringer} H., {Matsushita} K., {Churazov} E., {Ikebe} Y., {Chen} Y., Feb.
  2002, \aap, 382, 804

\bibitem[{{B{\"o}hringer} et~al.(2004){B{\"o}hringer}, {Matsushita},
  {Churazov}, {Finoguenov}, \& {Ikebe}}]{Boehringer04}
{B{\"o}hringer} H., {Matsushita} K., {Churazov} E., {Finoguenov} A., {Ikebe}
  Y., Mar. 2004, \aap, 416, L21

\bibitem[{{Br{\"u}ggen}(2002)}]{Brueggen02metals}
{Br{\"u}ggen} M., May 2002, \apjl, 571, L13

\bibitem[{{Br{\"u}ggen} \& {Kaiser}(2002)}]{Brueggen02}
{Br{\"u}ggen} M., {Kaiser} C.R., Jul. 2002, \nat, 418, 301

\bibitem[{{Cappellari} \& {Copin}(2003)}]{Cappellari03}
{Cappellari} M., {Copin} Y., Jun. 2003, \mnras, 342, 345

\bibitem[{{Cappellaro} et~al.(1999){Cappellaro}, {Evans}, \&
  {Turatto}}]{Cappellaro99}
{Cappellaro} E., {Evans} R., {Turatto} M., Nov. 1999, \aap, 351, 459

\bibitem[{{Churazov} et~al.(2001){Churazov}, {Br{\"u}ggen}, {Kaiser},
  {B{\"o}hringer}, \& {Forman}}]{Churazov01}
{Churazov} E., {Br{\"u}ggen} M., {Kaiser} C.R., {B{\"o}hringer} H., {Forman}
  W., Jun. 2001, \apj, 554, 261

\bibitem[{{Ciotti} et~al.(1991){Ciotti}, {Pellegrini}, {Renzini}, \&
  {D'Ercole}}]{Ciotti91}
{Ciotti} L., {Pellegrini} S., {Renzini} A., {D'Ercole} A., Aug. 1991, \apj,
  376, 380

\bibitem[{{Croton} et~al.(2006){Croton}, {Springel}, {White} et~al.}]{Croton06}
{Croton} D.J., {Springel} V., {White} S.D.M., et~al., Jan. 2006, \mnras, 365,
  11

\bibitem[{{De Luca} \& {Molendi}(2004)}]{DeLuca04}
{De Luca} A., {Molendi} S., Jun. 2004, \aap, 419, 837

\bibitem[{{de Plaa} et~al.(2006){de Plaa}, {Werner}, {Bykov} et~al.}]{dePlaa06}
{de Plaa} J., {Werner} N., {Bykov} A.M., et~al., Jun. 2006, \aap, 452, 397

\bibitem[{{Dennison}(1980)}]{dennison1980}
{Dennison} B., Mar. 1980, \apj, 236, 761

\bibitem[{{Diehl} \& {Statler}(2006)}]{Voronoi06}
{Diehl} S., {Statler} T.S., May 2006, \mnras, 368, 497

\bibitem[{{Feigelson} et~al.(1987){Feigelson}, {Wood}, {Schreier}, {Harris}, \&
  {Reid}}]{Feigelson87}
{Feigelson} E.D., {Wood} P.A.D., {Schreier} E.J., {Harris} D.E., {Reid} M.J.,
  Jan. 1987, \apj, 312, 101

\bibitem[{{Feretti} \& {Neumann}(2006)}]{feretti2006}
{Feretti} L., {Neumann} D.M., May 2006, \aap, 450, L21

\bibitem[{{Finoguenov} et~al.(2002){Finoguenov}, {Matsushita}, {B{\"o}hringer},
  {Ikebe}, \& {Arnaud}}]{Finoguenov02}
{Finoguenov} A., {Matsushita} K., {B{\"o}hringer} H., {Ikebe} Y., {Arnaud} M.,
  Jan. 2002, \aap, 381, 21

\bibitem[{{Forman} et~al.(2005){Forman}, {Nulsen}, {Heinz} et~al.}]{Forman05}
{Forman} W., {Nulsen} P., {Heinz} S., et~al., Dec. 2005, ApJ, 635, 894

\bibitem[{{Forman} et~al.(2007){Forman}, {Jones}, {Churazov} et~al.}]{Forman06}
{Forman} W., {Jones} C., {Churazov} E., et~al., Aug. 2007, \apj, 665, 1057

\bibitem[{{Fusco-Femiano} et~al.(2004){Fusco-Femiano}, {Orlandini}, {Brunetti}
  et~al.}]{fusco-femiano2004}
{Fusco-Femiano} R., {Orlandini} M., {Brunetti} G., et~al., Feb. 2004, \apjl,
  602, L73

\bibitem[{{Gastaldello} \& {Molendi}(2002)}]{Gastaldello02}
{Gastaldello} F., {Molendi} S., Jun. 2002, \apj, 572, 160

\bibitem[{{Goldshmidt} \& {Rephaeli}(1993)}]{goldshmidt1993}
{Goldshmidt} O., {Rephaeli} Y., Jul. 1993, \apj, 411, 518

\bibitem[{{Harms} et~al.(1994){Harms}, {Ford}, {Tsvetanov} et~al.}]{Harms94}
{Harms} R.J., {Ford} H.C., {Tsvetanov} Z.I., et~al., Nov. 1994, \apjl, 435, L35

\bibitem[{{Henry} et~al.(2004){Henry}, {Finoguenov}, \& {Briel}}]{Henry04}
{Henry} J.P., {Finoguenov} A., {Briel} U.G., Nov. 2004, \apj, 615, 181

\bibitem[{{Iwamoto} et~al.(1999){Iwamoto}, {Brachwitz}, {Nomoto}
  et~al.}]{Iwamoto99}
{Iwamoto} K., {Brachwitz} F., {Nomoto} K., et~al., Dec. 1999, \apjs, 125, 439

\bibitem[{{Kaastra} et~al.(1996){Kaastra}, {Mewe}, \& {Nieuwenhuijzen}}]{spex}
{Kaastra} J.S., {Mewe} R., {Nieuwenhuijzen} H., 1996, In: UV and X-ray
  Spectroscopy of Astrophysical and Laboratory Plasmas p.411, K. Yamashita and
  T. Watanabe. Tokyo : Universal Academy Press

\bibitem[{{Kaastra} et~al.(2004){Kaastra}, {Tamura}, {Peterson}
  et~al.}]{Kaastra04}
{Kaastra} J.S., {Tamura} T., {Peterson} J.R., et~al., Jan. 2004, \aap, 413, 415

\bibitem[{{Kobayashi} \& {Arimoto}(1999)}]{Kobayashi99}
{Kobayashi} C., {Arimoto} N., Dec. 1999, \apj, 527, 573

\bibitem[{{Lieu} et~al.(1996){Lieu}, {Mittaz}, {Bowyer} et~al.}]{Lieu96}
{Lieu} R., {Mittaz} J.P.D., {Bowyer} S., et~al., Feb. 1996, \apjl, 458, L5

\bibitem[{{Lieu} et~al.(1999){Lieu}, {Ip}, {Axford}, \& {Bonamente}}]{Lieu99}
{Lieu} R., {Ip} W.H., {Axford} W.I., {Bonamente} M., Jan. 1999, \apjl, 510, L25

\bibitem[{{Lodders}(2003)}]{Lodders}
{Lodders} K., Jul. 2003, \apj, 591, 1220

\bibitem[{{Mahdavi} et~al.(2005){Mahdavi}, {Finoguenov}, {B{\"o}hringer},
  {Geller}, \& {Henry}}]{Mahdavi05}
{Mahdavi} A., {Finoguenov} A., {B{\"o}hringer} H., {Geller} M.J., {Henry} J.P.,
  Mar. 2005, \apj, 622, 187

\bibitem[{{Matsushita} et~al.(2002){Matsushita}, {Belsole}, {Finoguenov}, \&
  {B{\"o}hringer}}]{Matsushita02}
{Matsushita} K., {Belsole} E., {Finoguenov} A., {B{\"o}hringer} H., Apr. 2002,
  \aap, 386, 77

\bibitem[{{Matsushita} et~al.(2003){Matsushita}, {Finoguenov}, \&
  {B{\"o}hringer}}]{Matsushita03}
{Matsushita} K., {Finoguenov} A., {B{\"o}hringer} H., Apr. 2003, \aap, 401, 443

\bibitem[{{Matsushita} et~al.(2007){Matsushita}, {B{\"o}hringer}, {Takahashi},
  \& {Ikebe}}]{Matsushita07}
{Matsushita} K., {B{\"o}hringer} H., {Takahashi} I., {Ikebe} Y., Feb. 2007,
  \aap, 462, 953

\bibitem[{{McNamara} \& {Nulsen}(2007)}]{McNamara07}
{McNamara} B.R., {Nulsen} P.E.J., Sep. 2007, \araa, 45, 117

\bibitem[{{Molendi}(2002)}]{Molendi02}
{Molendi} S., Dec. 2002, \apj, 580, 815

\bibitem[{{Omma} et~al.(2004){Omma}, {Binney}, {Bryan}, \& {Slyz}}]{Omma04}
{Omma} H., {Binney} J., {Bryan} G., {Slyz} A., Mar. 2004, \mnras, 348, 1105

\bibitem[{{Owen} et~al.(2000){Owen}, {Eilek}, \& {Kassim}}]{OEK00}
{Owen} F.N., {Eilek} J.A., {Kassim} N.E., Nov. 2000, \apj, 543, 611

\bibitem[{{Peletier} et~al.(1990){Peletier}, {Davies}, {Illingworth}, {Davis},
  \& {Cawson}}]{Peletier90}
{Peletier} R.F., {Davies} R.L., {Illingworth} G.D., {Davis} L.E., {Cawson} M.,
  Oct. 1990, \aj, 100, 1091

\bibitem[{{Peterson} \& {Fabian}(2006)}]{Peterson06}
{Peterson} J.R., {Fabian} A.C., Apr. 2006, \physrep, 427, 1

\bibitem[{{Rafferty} et~al.(2006){Rafferty}, {McNamara}, {Nulsen}, \&
  {Wise}}]{Rafferty06}
{Rafferty} D.A., {McNamara} B.R., {Nulsen} P.E.J., {Wise} M.W., Nov. 2006,
  \apj, 652, 216

\bibitem[{{Read} \& {Ponman}(2003)}]{ReadPonman}
{Read} A.M., {Ponman} T.J., Oct. 2003, \aap, 409, 395

\bibitem[{{Rebusco} et~al.(2006){Rebusco}, {Churazov}, {B{\"o}hringer}, \&
  {Forman}}]{Rebusco06}
{Rebusco} P., {Churazov} E., {B{\"o}hringer} H., {Forman} W., Nov. 2006,
  \mnras, 372, 1840

\bibitem[{{Roediger} et~al.(2007){Roediger}, {Br{\"u}ggen}, {Rebusco},
  {B{\"o}hringer}, \& {Churazov}}]{Roediger06}
{Roediger} E., {Br{\"u}ggen} M., {Rebusco} P., {B{\"o}hringer} H., {Churazov}
  E., Feb. 2007, \mnras, 375, 15

\bibitem[{{Salpeter}(1955)}]{Salpeter55}
{Salpeter} E.E., Jan. 1955, \apj, 121, 161

\bibitem[{{Sanders} et~al.(2005){Sanders}, {Fabian}, \& {Dunn}}]{sanders2005}
{Sanders} J.S., {Fabian} A.C., {Dunn} R.J.H., Jun. 2005, \mnras, 360, 133

\bibitem[{{Sarazin}(1986)}]{Sarazin86}
{Sarazin} C.L., Jan. 1986, Reviews of Modern Physics, 58, 1

\bibitem[{{Sijacki} et~al.(2007){Sijacki}, {Springel}, {di Matteo}, \&
  {Hernquist}}]{Sijacki07}
{Sijacki} D., {Springel} V., {di Matteo} T., {Hernquist} L., Sep. 2007, \mnras,
  380, 877

\bibitem[{{Simionescu} et~al.(2007){Simionescu}, {B{\"o}hringer},
  {Br{\"u}ggen}, \& {Finoguenov}}]{Simionescu07}
{Simionescu} A., {B{\"o}hringer} H., {Br{\"u}ggen} M., {Finoguenov} A., Apr.
  2007, \aap, 465, 749

\bibitem[{{Spitzer}(1962)}]{Spitzer62}
{Spitzer} L., 1962, {Physics of Fully Ionized Gases}, Physics of Fully Ionized
  Gases, New York: Interscience (2nd edition), 1962

\bibitem[{{Tonry} et~al.(2001){Tonry}, {Dressler}, {Blakeslee}
  et~al.}]{Tonry01}
{Tonry} J.L., {Dressler} A., {Blakeslee} J.P., et~al., Jan. 2001, \apj, 546,
  681

\bibitem[{{Watson} et~al.(2001){Watson}, {Augu{\`e}res}, {Ballet}
  et~al.}]{Watson01}
{Watson} M.G., {Augu{\`e}res} J.L., {Ballet} J., et~al., Jan. 2001, \aap, 365,
  L51

\bibitem[{{Werner} et~al.(2006{\natexlab{a}}){Werner}, {B{\"o}hringer},
  {Kaastra} et~al.}]{Werner06}
{Werner} N., {B{\"o}hringer} H., {Kaastra} J.S., et~al., Nov.
  2006{\natexlab{a}}, \aap, 459, 353

\bibitem[{{Werner} et~al.(2006{\natexlab{b}}){Werner}, {de Plaa}, {Kaastra}
  et~al.}]{Werner06_2A0335}
{Werner} N., {de Plaa} J., {Kaastra} J.S., et~al., Apr. 2006{\natexlab{b}},
  \aap, 449, 475

\bibitem[{{Young} et~al.(2002){Young}, {Wilson}, \& {Mundell}}]{Young02}
{Young} A.J., {Wilson} A.S., {Mundell} C.G., Nov. 2002, \apj, 579, 560

\end{thebibliography}

\end{document}